 \documentclass[twocol]{ametsoc}  

 \journal{jas}

\bibpunct{(}{)}{;}{a}{}{,}

\title{
Constraints from invariant subtropical vertical velocities on the scalings\\of Hadley cell strength and downdraft width with rotation rate}

\authors{Jonathan L. Mitchell\correspondingauthor{Jonathan L. Mitchell, 595 Charles Young Drive E., Los Angeles, CA 90095.}}
\affiliation{Department of Atmospheric and Oceanic Sciences and Department of Earth, Planetary, and Space Sciences, UCLA, Los Angeles, CA}
\email{jonmitch@ucla.edu}

\extraauthor{Spencer A. Hill}
\extraaffil{Lamont-Doherty Earth Observatory, Columbia University, Palisades, NY}

\abstract{Weak-temperature-gradient influences from the tropics and quasigeostrophic influences from the extratropics plausibly constrain the subtropical-mean static stability in terrestrial atmospheres.  Because mean descent acting on this static stability is a leading-order term in the thermodynamic balance, a state-invariant static stability would impose constraints on the Hadley cells, which this paper explores in simulations of varying planetary rotation rate.  If downdraft-averaged effective heating (the sum of diabatic heating and eddy heat flux convergence) too is invariant, so must be vertical velocity --- an ``omega governor.''  In that case, the Hadley circulation overturning strength and downdraft width must scale identically --- the cell can strengthen only by widening or weaken only by narrowing.  Simulations in two idealized, dry GCMs with a wide range of planetary rotation rates exhibit nearly unchanging downdraft-averaged static stability, effective heating, and vertical velocity, as well as nearly identical scalings of the Hadley cell downdraft width and strength.  In one, eddy stresses set this scaling directly (the Rossby number remains small); in the other, eddy stress and bulk Rossby number changes compensate to yield the same, \({\sim}\Omega^{-1/3}\) scaling.  The consistency of this power law for cell width and strength variations may indicate a common driver, and we speculate that Ekman pumping could be the mechanism responsible for this behavior.  Extending to moist atmospheres, in an idealized aquaplanet GCM the subtropical static stability is also insensitive to rotation rate but the effective heating and vertical velocity are not.
}

\begin{document}

\maketitle

\section{Introduction}

Terrestrial atmospheres in which both tropical and extratropical dynamical regimes exist are thought to be common \citep[as on Earth, Mars, and likely many identified exoplanets; e.g.][]{showman_atmospheric_2014}.  Their atmospheres are neither dominated by zonally banded dynamics as on the gas giants, nor by global Hadley cells as on Venus and Titan.  Instead, the Hadley cell descending branches sit roughly in the transitional, subtropical zone.  In this planetary context, ``subtropical'' is not restricted to the \(\sim 25-35^\circ\) latitude band corresponding to the subtropics for Earth, but rather wherever the transitional region falls in which the large-scale circulation is neither strongly divergent and convectively driven as in the deep Tropics nor sufficiently horizontal so as to be unambiguously quasigeostrophic as in the extratropics.

This inherent intermediacy of the subtropics provides the Hadley cell descending branch a certain freedom that limits the predictive power of existing dynamical theories based on either limit of the zonal momentum balance, usually expressed in terms of the local Rossby number ($Ro_L$, defined formally below).  On the one hand, nearly inviscid, axisymmetric, \({Ro_L\sim1}\) theories \citep{Held_Hou80} that are reasonable and useful for the deep tropics neglect the often leading-order zonally asymmetric eddy processes \citep[e.g.][]{Becker_etal97}.  On the other, the eddy-driven, \(Ro_L\ll1\) theories \citep[e.g.][]{Walker_Schneider06} entirely neglect
vorticity advection by the mean flow, an assumption at odds with observed and simulated subtropical angular momentum distributions \citep[e.g.][]{Schneider_06}.  Moreover, even if a particular Hadley cell happens to fall squarely within either regime, this is merely diagnostic: in response to some perturbation it can move between these limits (increase or decrease its $Ro_L$ value) just as well as it can respond as the relevant limiting theory would predict.

Regardless of the dynamical regime, the adiabatic warming generated by mean descent in the Hadley cell downdraft acting on positive static stability is virtually always of leading-order thermodynamic importance, balancing the combination of diabatic cooling and eddy heat flux divergence or convergence.  And unlike the variations in the dynamical regime, there are good reasons to expect the static stability to vary only modestly across planetary parameters for a given diabatic forcing distribution.  From the deep tropics side, the static stability is constrained by convection and wave dynamics to be roughly constant and nearly independent of planetary rotation rate \citep{Sobel_etal01}.  From the extratropical side, quasigeostrophic arguments \citep{jansen_equilibration_2013} suggest that the extratropical static stability will stay fixed as planetary rotation rate is varied.  Given these influences emanating from both poleward and equatorward sides, it is reasonable to imagine that the subtropical static stability (or at least its average) is likewise constrained.\footnote{Our working definition of the downdraft is the latitudinal extent from the streamfunction maximum to the latitude where it drops to 10\% of the maximum going poleward.  Downdraft-averaged quantities are defined as the meridional average across these latitudes at the level of the maximum of the streamfunction.}

If, in addition to the static stability, the meridional average over the downdraft at the level of the maximum streamfunction of the diabatic heating plus eddy heat flux divergence, $Q_\mathrm{eff}$ (formally defined below), is independent of rotation rate, then the magnitude of vertical velocities must be fixed.  We term this the ``omega governor'' because vertical (pressure) velocities are limited or ``governed'' by fixed static stability and $Q_{\rm{eff}}$.  Even more so than for the static stability, there's no \emph{a-priori} argument for constant $Q_{\rm{eff}}$.  Yet we will show that both the downdraft-averaged static stability and eddy-plus-diabatic heating terms --- and with them, vertical velocity --- \emph{do} stay fixed across simulations in two idealized, dry general circulation models.

By ``dry'', we refer to the absence of an explicit representation of the effects of latent heating.  We do, however, implicitly include the stabilizing effect of latent heating in the temperature relaxation or adjustment by processes described in Section \ref{sec:experiments}.  In this sense, one might interpret our dry-stable atmospheres as being representative of ``moist'' atmospheres that lack latent effects associated with dynamical moisture transport and convergence.  This difference appears to be quite important, as described in Section \ref{sec:moist}.  Thus, our working definition of ``dry'' model forcing is shorthand for ``dry-stable''.  Note that even in dry simulations that are adjusted to dry-neutral stability by convection, the circulation itself creates a stable region by advecting high-entropy air into the upper troposphere from the stratosphere \citep{Caballero_etal_08}.

The implications of the omega governor for Hadley cells are this paper's focus.  We show that the omega governor constraint closely links eddy stresses to the strength \emph{and} width of the Hadley cell downdraft; formally, we develop a new theory that yields identical scalings with rotation rate for the width of the Hadley cell downdraft and the Hadley cell overturning strength.  In Section \ref{sec:theory} we present the theory.  We then test it against simulations in two dry, idealized GCMs; the model and simulation specifications are described in Section~\ref{sec:experiments} and the results in Section~\ref{sec:results}.  Section~\ref{sec:moist} presents the results of analogous simulations performed in an idealized, moist, aquaplanet GCM.  We conclude in Section~\ref{sec:disc} with a summary and discussion.

\section{Theory}
\label{sec:theory}

\subsection{Why downdraft-averaged \(\omega\) should be insensitive to \(\Omega\)}
In the deep tropics, convection within the Intertropical Convergence Zone (ITCZ) controls the local static stability \citep[e.g.][]{Sobel_etal01}.  The small Coriolis parameter then communicates this concentrated diabatic heating horizontally via gravity waves \citep{satoh_hadley_1994,fang_simple_1996}.  This Weak Temperature Gradient (WTG) constraint \citep{Sobel_etal01} can be expected to have some influence into the subtropics, even though the Coriolis parameter grows and WTG is less valid moving poleward.  This applies equally to dry atmospheres as to moist atmospheres, with the presence of moist convection (or its organization into single or double ITCZ structures) primarily altering the static stability itself \citep[e.g.][]{numaguti_dynamics_1993,blackburn_aqua-planet_2013}, rather than the meridional extent of the WTG zone. 

In the extratropics, to first order the large-scale dynamics are quasigeostrophic, and the predominant heat balance is between the residual mean overturning and eddy flux divergences.  In dry atmospheres, the extratropical eddy heat fluxes are primarily adiabatic, i.e.~along isentropes, and in marginally or strongly baroclinically supercritical atmospheres (which should characterize the extratropics of most terrestrial atmospheres), this causes the extratropical static stability to vary weakly with planetary rotation rate \citep{jansen_equilibration_2013}.

Combining these tropical and extratropical influences, then, to first order one can expect a meridional profile of static stability in the free troposphere that is flat both at low latitudes and (though not generally at the same value as in the Tropics) in mid-latitudes.  This means the static stability will vary in latitude in the subtropics from the tropical to extratropical values.  But importantly the \emph{average} subtropical value is unlikely to change even if the width of the subtropics changes, as is likely under rotation rate variations.  In the simplest case of a linear variation of static stability with latitude over the subtropics, the average value will remain fixed. 

There is less theoretical basis for $Q_{\rm{eff}}$ averaged over the Hadley cell downdraft to be independent of rotation rate.  Particularly for moist atmospheres but in dry atmospheres as well, the diabatic heating will depend on the circulation itself as well as (in models) the physical parameterizations of radiation and convection.  In addition, breaking eddies generally act to transport heat poleward, overlapping with the heat transport of the Hadley cell but extending further into the extratropics \citep{Trenberth_Stepaniak03}.  In principle, this alters the thermodynamic constraint on vertical velocities in the downdraft, but in practice the eddy heat flux convergence can simply be treated as a separate heating term contributing to the zonal-mean energetics.\footnote{There is a rich history of treating the eddy heat (and momentum) flux divergences diagnostically as forcing terms on the mean meridional circulation, perhaps most formally via the Kuo-Eliassen equation \citep{eliassen_slow_1951,kuo_forced_1956,lorenz_nature_1967}, which remains in use in recent times \citep[e.g.][]{chemke_ocean_2018}, or by imposing diagnosed eddy stress and heating fields as forcings in quasi-axisymmetric simulations \citep[e.g.\@][and references therein]{Singh_Kuang16,singh_eddy_2016}.}

Moreover, as shown below the sum of diabatic and eddy heat flux convergence, $Q_\mathrm{eff}$, is remarkably constant across a broad range of rotation rates in both dry models we use. We thus take an invariant subtropical-mean effective heating (at least for planetary rotation rate variations) as an empirically justified result for dry atmospheres, whose theoretical underpinnings remain to be determined.

With both the static stability and effective heating fixed in the downdraft average, then provided their covariance within the downdraft also varies weakly, it follows that the descent rate averaged over the Hadley cell downdraft must also be fixed.  More formally, begin with the time-mean, zonal-mean thermodynamic equation in regions of purely vertical zonal-mean flow such that horizontal advection by the mean flow vanishes:
\begin{equation}\label{eq:zonalthermo}
\bar\omega \frac{\partial\bar\theta}{\partial p} + \frac{1}{a\cos\varphi}\frac{\partial}{\partial\varphi}(\cos\overline{v'\theta'}) + \frac{\partial}{\partial p} (\overline{\omega'\theta'})= \bar{Q}\left(\frac{p}{p_s}\right)^{-\kappa} 
\end{equation}
with potential temperature $\theta$, $\kappa = R/C_p$, gas constant $R$, specific heat at constant pressure $C_p$, diabatic forcing $\bar{Q}$ in K/s, overbars denoting temporal and zonal averages, and primes deviations therefrom.  The divergence of eddy heat fluxes are represented by the second and third terms on the LHS of (\ref{eq:zonalthermo}).  Solving (\ref{eq:zonalthermo}) for zonal-mean vertical velocity and taking the meridional mean over the downdraft,
\begin{equation}\label{eq:omegathermo}
\left[\bar\omega\right] = \left[\frac{Q_\mathrm{eff}}{\frac{\partial\bar\theta}{\partial p}}\right] ,
\end{equation}
where square brackets constitute meridional averages spanning the downdraft of the Hadley cell and
\begin{equation}\label{eq:Ro0HR}
Q_\mathrm{eff} \equiv Q_{\rm tot} + {\rm EHFC} ,
\end{equation}
where $Q_\mathrm{tot} \equiv \bar{Q}\left(\frac{p}{p_s}\right)^{-\kappa}$ and
${\rm EHFC} \equiv - \frac{1}{a\cos\varphi}\frac{\partial}{\partial\varphi}(\cos\overline{v'\theta'}) - \frac{\partial}{\partial p} (\overline{\omega'\theta'})$.
The ``eff'' subscript underscores that this is an ``effective'' heating experienced by the mean circulation, including as it does both the diabatic heating and the eddy heat flux convergence.  In the strongly eddying, $Ro\ll1$ limit, the eddy terms in (\ref{eq:Ro0HR}) are likely important.\footnote{Though, strictly, they need not be; in principle eddies (especially barotropic ones) could effect large stresses but small eddy heat flux divergences.}  In the axisymmetric, $Ro\sim1$ limit, (\ref{eq:Ro0HR}) reduces to the diabatic heating provided transient symmetric instabilities do not generate appreciable heat flux divergences.  In non-eddy-permitting simulations, the response of the Hadley cell to imposed EHFC-like heating is primarily angular-momentum-conserving \citep{singh_eddy_2016}, as would be expected from thermodynamic considerations, and lends credibility to our grouping of EHFC and $Q_\mathrm{tot}$ into an effective heating, $Q_\mathrm{eff}$.

With \([Q_\mathrm{eff}]\) and \([\partial\theta/\partial p]\) fixed, (\ref{eq:omegathermo}) implies a value for $\left[\bar\omega\right]$ that is independent of rotation rate --- and, importantly, of the Rossby number.

\subsection{Hadley cell dynamical theories}
If an atmosphere is axisymmetric, inviscid above the boundary layer, and with ascent out of the boundary layer occurring at a single latitude, the resulting free-tropospheric flow must be angular-momentum-conserving (AMC). The local Rossby number ($Ro_L=-\overline{\zeta}/f$, where $\overline\zeta=-\partial\overline{u}/\partial y$ is the zonal-mean relative vorticity, and $f$ is the Coriolis parameter) then is unity throughout the Hadley cells \citep{Held_Hou80}.  Axisymmetric theory compares the vertically integrated thermodynamic equation in radiative-convective equilibrium to its gradient-balanced AMC counterpart and then appeals to temperature continuity and energy conservation to predict the Hadley cell width.  Notably, the AMC perspective primarily constrains the width of the Hadley circulation; one must appeal to the thermodynamic balance of the descending branch described above to estimate the circulation strength, typically measured by the maximum of the overturning streamfunction.  The AMC model requires that eddy stresses and heat flux divergences be negligible throughout the Hadley cells.  The resulting dependence of the Hadley cell width and strength on planetary rotation are quite steep: $\Omega^{-1}$ and $\Omega^{-3}$, respectively.

Eddy stresses, however, are not generally negligible, and angular momentum is not generally uniform even along individual streamlines, let alone over the expanse of the entire circulation \citep[e.g.\@][]{adam_global_2009,Hill_etal19}.  Baroclinic eddies generated in mid-latitudes propagate equatorward and break, drawing westerly momentum poleward out of the subtropics.  In response, the Hadley circulation transports angular momentum poleward to meet the momentum demand of the breaking eddies.  In this way and in the limit $Ro_L\ll 1$, the strength of the Hadley circulation can be thought of as being determined by the momentum budget in the subtropics \cite[e.g.,][]{Becker_etal97, Kim_Lee01, Walker_Schneider06, held_2019}.

Neither limit is perfect --- large portions of the tropical free troposphere on Earth and other atmospheres are routinely in an intermediate regime, $Ro<1$ but not $Ro\ll 1$ --- but there are times/locations where $Ro\geq0.7$, mostly in the deep Tropics, while in the subtropics often $Ro\sim0.1$ \citep{Schneider_06}.

\subsection{Combining fixed \([\overline{\omega}]\) with existing Hadley cell theory: the ``omega governor''}
We define the Hadley cell strength as the maximum within the cell of the zonal-mean mass streamfunction,
\begin{equation}
\Psi(\varphi,p) = 2\pi a\cos\varphi\int_{p_s}^p \bar{v} \frac{dp}{ g} \ ,
\end{equation}
with planetary radius $a$, latitude $\varphi$, surface pressure $p_s$, meridional velocity $v$, gravity $g$, and an overbar denoting a temporal and zonal average.  Equivalently, the streamfunction can be calculated as
\begin{equation} \label{eq:psimaxomega}
\Psi(\varphi,p) = - \frac{2\pi a^2}{g}\int_{\varphi_0}^{\varphi} \bar\omega\cos\varphi d\varphi \ ,
\end{equation}
with zonal- and time-mean vertical (pressure) velocity $\bar\omega$ and $\varphi_0$ being a latitude where the streamfunction is zero, i.e. at the cell edges.

In the upper, poleward-flowing branch of the Hadley circulation near the latitude of the maximum mass flux (i.e. the cell center), we expect a dominant balance between meridional advection of absolute vorticity and eddy momentum flux divergence,
\begin{equation}\label{eq:momentum}
(f+\bar{\zeta})\bar{v} = (1-Ro_L)f\bar{v}\simeq s \ ,
\end{equation}
where
\begin{equation}
s = \frac{1}{a\cos\varphi}\frac{\partial}{\partial\varphi}(\cos\varphi\overline{u'v'})+\frac{\partial}{\partial p}(\overline{u'\omega'})
\end{equation}
is the eddy momentum flux divergence, $\bar{\zeta}$ is the zonal mean relative vorticity, $f$ is the planetary vorticity, and $u'$ and $v'$ are the deviations from the zonal and time mean of the horizontal wind \citep{Walker_Schneider06}.  If $Ro_L= 1$, it must be the case that $s=0$, and thus (\ref{eq:momentum}) is degenerate, providing no constraint on the flow.  Integrating (\ref{eq:momentum}) vertically from the level of the streamfunction maximum, $p_m$, to the top of the circulation where there is vanishing mass flux, $p_t$,
\begin{equation}
\int_{p_m}^{p_t}(1-Ro_L)f\bar{v} \frac{dp}{g} \simeq \int_{p_m}^{p_t}s \frac{dp}{g} \ .
\end{equation}
Then multiplying both sides by $2\pi a \cos\varphi$, and, following \cite{Singh_Kuang16}, defining a mass-weighted bulk Rossby number\footnote{$Ro_L$ and $Ro$ are to be distinguished from the thermal Rossby number, $Ro_{\rm th}$, which is a control parameter in the theory of \cite{Held_Hou80} and others.  The thermal Rossby number is a scalar defined in terms of planetary parameters with a specified dependence on rotation rate, $Ro_{\rm th}\sim\Omega^{-2}$, while the local ($Ro_L$) and bulk ($Ro$) Rossby numbers are functions of latitude and pressure and are diagnosed from a given simulation.}, $Ro$, we arrive at
\begin{equation}\label{eq:psimax}
\Psi_{\rm vmax}(1-Ro) \simeq S/f
\end{equation}
where $S = 2\pi a \cos\varphi \int_{p_m}^{p_t}s \ dp/g$ and $\Psi_{\rm vmax}(\varphi)$ is the value of the streamfunction at the level of the overall Hadley cell maximum mass flux.\footnote{Note that our (\ref{eq:psimax}) differs from that in \cite{Singh_Kuang16} in that the mass streamfunction is in kg/s rather than kg m$^{-1}$ s$^{-1}$.},\footnote{While (\ref{eq:psimax}) is valid at any latitude meeting the given assumptions, henceforth we use it only at the latitude of the maximum streamfunction, $\varphi=\varphi_{\rm max}$.}
(\ref{eq:psimax}) makes clear the dependence of the max streamfunction on the (bulk) Rossby number: if $Ro\ll1$, the demand for momentum by breaking extratropical eddies (RHS of (\ref{eq:psimax})) must be consistent with poleward Hadley cell momentum transport; if $Ro\sim1$, the momentum budget does not constrain the Hadley circulation strength.

More formally, since (\ref{eq:psimaxomega}) and (\ref{eq:psimax}) must equal one another at the level of the maximum mass flux of the Hadley cell, $p_m$, we have
\begin{equation}\label{eq:full2psi}
S/f \simeq -(1-Ro)\frac{2\pi a^2\left[\bar{\omega}\right]_{\rm vmax}}{g} (\sin\varphi - \sin\varphi_0)  \ ,
\end{equation}
with zonal- and meridional-mean vertical velocity at the level of the maximum mass flux,
\begin{equation}
\left[\bar{\omega}\right]_{\rm vmax} = \frac{\int_{\varphi_0}^{\varphi} \bar{\omega}(\varphi,p_m)\cos\varphi d\varphi}{ \sin\varphi-\sin\varphi_0} \ .
\end{equation}
Setting $\varphi=\varphi_\mathrm{max}$ and $\varphi_0=\varphi_h$, where $\varphi_h$ is the cell's poleward edge, (\ref{eq:full2psi}) becomes
\begin{equation}\label{eq:2psi}
S/f \simeq -\frac{2\pi a^2\left[\bar{\omega}\right]_{\rm vmax}}{g} (\sin\varphi_{\rm max} - \sin\varphi_h) \ {\rm if} \ Ro\ll1 \ .
\end{equation}
If $\bar\omega_{\rm vmax}$ is constant and in the small-angle limit [i.e. $(\varphi_h, \varphi_{\rm max})\ll 1$], (\ref{eq:2psi}) further reduces to
\begin{equation}\label{eq:smallanglepsi}
S/f \simeq  \frac{2 \pi a^2\bar\omega_{\rm vmax}}{g}(\varphi_h-\varphi_{\rm max}) \ {\rm if} \ \  (Ro, \varphi_h, \varphi_{\rm max})\ll1 \ \ .
\end{equation}
By (\ref{eq:psimax}) and (\ref{eq:smallanglepsi}), respectively, the cell overturning strength and downdraft width (in latitude), $\varphi_h-\varphi_\mathrm{max}$, are both proportional to $S/f$.  Eddy stresses determine the overturning strength, which for a fixed ascent rate can only be altered as the planetary rotation rate is varied by making the cell downdraft narrower or wider \citep[see also][]{seo_mechanism_2014}.

This is not a closed theory, since both $S/f$ and $Ro$ must be diagnosed from simulations.  But given $S/f$ and $Ro$ the width and the strength have the same scaling with rotation rate.  If in addition $Ro$ is small, these scalings are set by --- and are identical to --- that of the eddy momentum flux divergence.  Indeed, the Hadley cell widths and strengths do follow simple and nearly identical power laws across a wide range of rotation rates in our dry simulations, in one model despite large variations in $Ro$, as we now describe.

\section{Model and simulation descriptions}
\label{sec:experiments}

We examine the accuracy of the preceding theoretical arguments using model simulations performed in two idealized GCMs with no latent heating.

The first model uses the Flexible Modeling System (FMS) spectral dynamical core \citep{Gordon_Stern82} forced with simple, linearized diabatic and frictional terms as in the Held-Suarez benchmark \citep{Held_Suarez94}.  The equilibrium temperature profile to which temperatures are relaxed via Newtonian cooling, from \citet{Mitchell_Vallis10}, differs from the original Held-Suarez benchmark in its vertical structure; in our model, we specify a uniform lapse rate of 6 K km$^{-1}$ to approximate a moist adiabat in the troposphere, which is capped by an isothermal stratosphere.  The horizontal structure of the forcing profile is the same as Held-Suarez, following a specified surface temperature meridional distribution $T_o = \bar{T}[1+\Delta_H/3(1-3\sin^2\varphi)]$, where $\bar{T}=285$K is the global-mean surface temperature and $\Delta_H=0.2$ is a non-dimensional equator-to-pole temperature gradient.  The stratospheric cap is specified by not allowing temperatures to drop below 70\% of $\bar{T}$.  The Newtonian cooling timescale distribution is identical to Held-Suarez.

The second model is an idealized GCM with convective relaxation that has been used in a number of prior studies investigating the general circulation \citep{Schneider_04, Walker_Schneider05, Walker_Schneider06, Schneider_Bordoni08, BordoniSchneider_10}.  The primary differences between this GCM and the Held-Suarez-like model are, first, the Newtonian cooling equilibrium temperature profile is statically unstable, and, second, it includes a simple convective scheme that relaxes temperatures over a uniform 4-hour timescale toward a lapse rate of $\gamma\Gamma_d$, where $\Gamma_\mathrm{d}$ is the dry adiabatic lapse rate and $\gamma$ = 0.7 mimics the stabilizing effect of condensation (although, again, the model is dry).  Also, rather than linear Rayleigh drag, this model uses a quadratic drag formulation within the planetary boundary layer (which has a top of $\sigma=0.85$ rather than at $\sigma=0.7$ as in the Held-Suarez case).  See \citet{Schneider_04} for further details.

Simulations are run over a broad range of planetary rotation rates, $\Omega^*=[1/16,1/8,1/4,1/2,1,2]$, where $\Omega^*=\Omega/\Omega_e$ and $\Omega_e$ is Earth's rotation rate.  Both models are run at T42 horizontal resolution for rotation rates smaller than that of Earth and at T85 for those with larger rotation rates, because the horizontal scale of circulations contract with increasing rotation rate.  Both models use 20 sigma-coordinate levels in the vertical, spaced evenly in the Held-Suarez case and unevenly in the convective adjustment case.  Averages are taken over the final 360 days of 1440 simulated days, with 6-hourly snapshots used for calculating eddy fields.

We perform the following data processing on all simulations: (1) quantities are symmetrized about the equator; (2) the Hadley cell strength, $\Psi_{\rm max}$, is taken as the maximum over the cell of the overturning mass flux; (3) the quantity $S/f$ is also taken at the latitude of $\Psi_{\rm max}$; and (4) vertical velocities are averaged over the downdraft at the level of maximum Hadley cell mass flux.  The Hadley cell edge is defined conventionally as the latitude where, moving poleward from the latitude and sigma level of the cell center, the mass flux drops to 10\% of the cell's maximum value.  The top of the Hadley cell is similarly defined, moving upward rather than poleward.

\section{Simulation results}
\label{sec:results}

\subsection{General features of large-scale circulation}
Before investigating the omega-governor-related fields in detail, we first summarize the overall character of the large-scale circulation across rotation rates and models.

\begin{figure*}
\includegraphics[width=\textwidth]{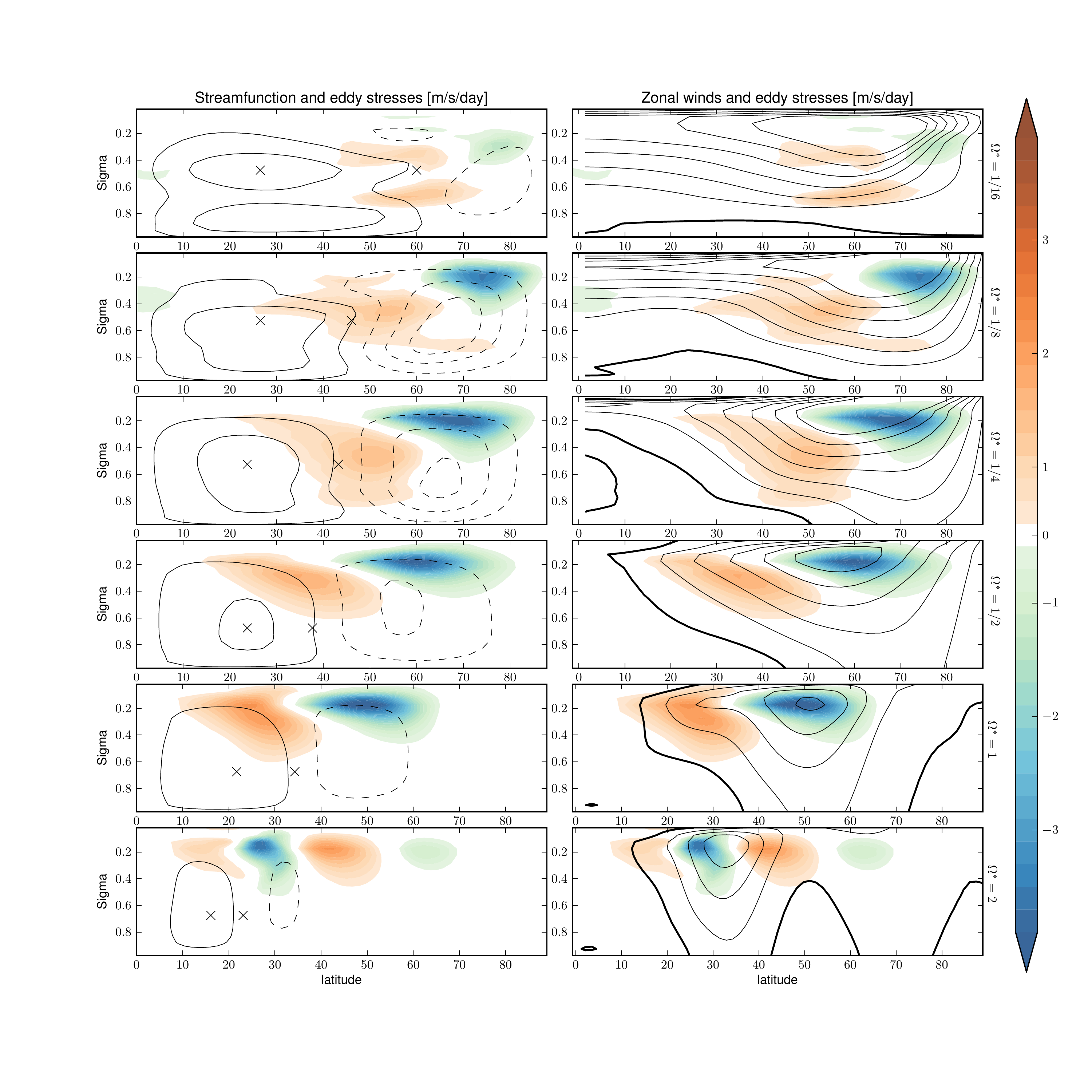}
\caption{Hemispherically symmetrized, Held-Suarez-like-forced model diagnostics, zonally and temporally averaged over the last 360 days of 1440-day simulations.  Values of $\Omega^*$ are labeled to the right.  Eddy momentum flux divergences weighted by $\cos\varphi$ (colors, both columns [m s$^{-1}$ day$^{-1}$]), mass streamfunctions (black contours, left column, positive values are solid and rotating clockwise, spaced -3$\times 10^{11}$ to 3$\times 10^{11}$ by 4$\times 10^10$), and zonal winds (right column, black contours spaced 0 to 50 by 8 m  s$^{-1}$, bolded contour is the zero-wind line).  The left ``x'' marks the location of the maximum mass flux of the Hadley cell and the right ``x'' marks our definition of the Hadley cell edge.
}
\label{fig:HS}
\end{figure*}

In the Held-Suarez model, the Hadley circulation generally widens and strengthens as the rotation rate decreases, as shown in the left column of Figure~\ref{fig:HS} (black contours).  Eddy momentum flux divergence (EMFD; colors) is evident in the upper, poleward quadrant of the Hadley cell at all rotation rates, marking the so-called ''surf zone'' where extratropical baroclinic eddies break and decelerate zonal winds.  EMFD in the extratropics drives a Ferell cell that also expands poleward and strengthens as rotation rate decreases, except for the slowest rotating case, $\Omega^*=1/16$.  Zonal winds (black contours, right column) generally follow the widening of the Hadley cell, moving poleward and strengthening with decreasing rotation rate.  At twice Earth's rotation rate, alternating patterns of positive and negative EMFD in latitude are apparent, and there is a hint of a multiple-jet configuration.  Simulations at 4 times Earth's rotation (not shown) exhibit multiple jets and meridional circulations, and those at rotation rates below $\Omega^*=1/16$ have effectively global Hadley cells (not shown); both regimes are beyond this paper's scope \citep[see e.g.][]{Williamsrangeunityplanetary1982,showman_atmospheric_2014}.

\begin{figure*}
\includegraphics[width=\textwidth]{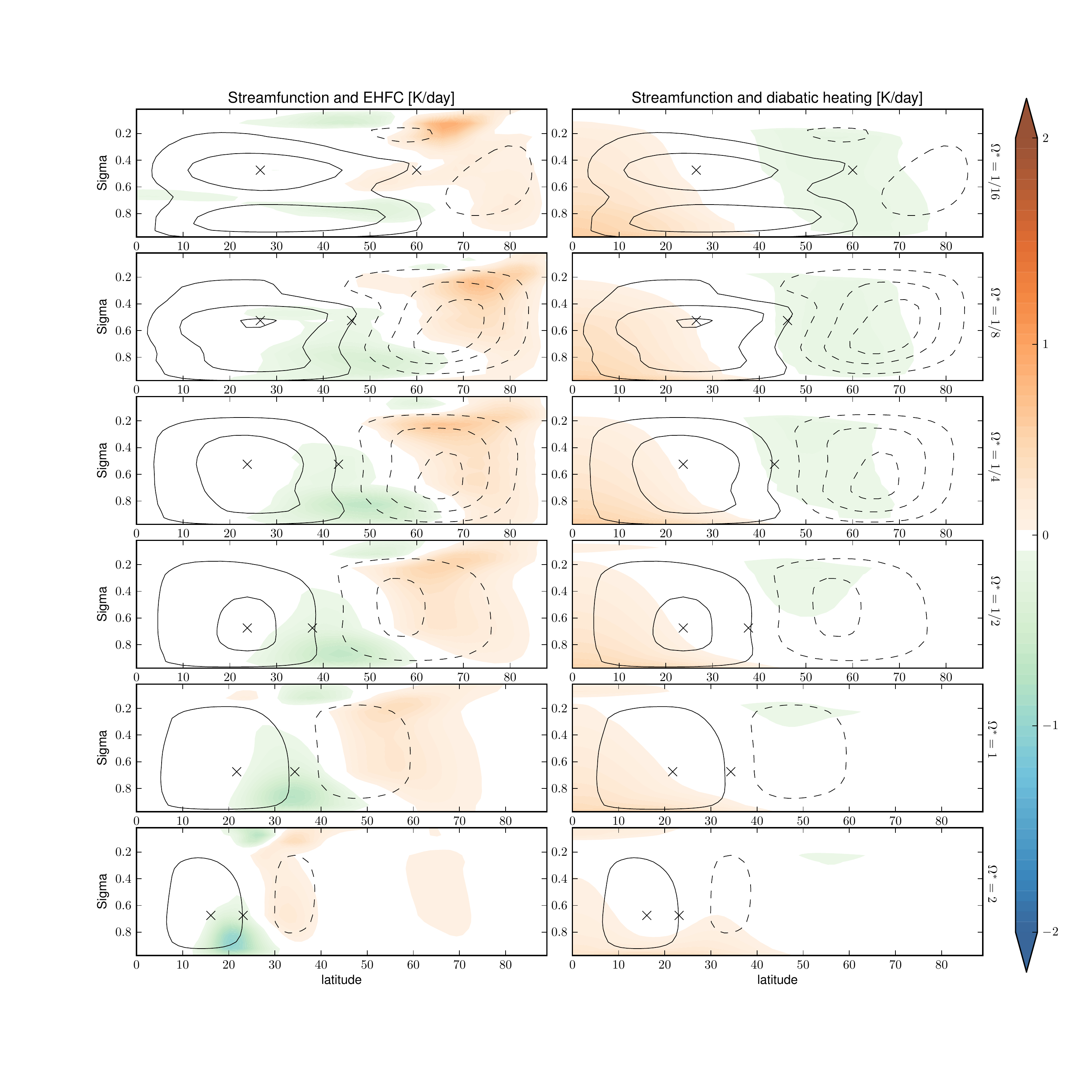}
\caption{(left) Eddy heat flux convergence weighted by $\cos\varphi$ [K day$^{-1}$] and mass streamfunction (as in Fig.~\ref{fig:HS}) for the Held-Suarez-like model forcing.  (right) Diabatic heating rate weighted by $\cos\varphi$ [K day$^{-1}$] and mass streamfunction (as in Fig.~\ref{fig:HS}). Values of $\Omega^*$ are shown along the right of each row.
}
\label{fig:HSheat}
\end{figure*}

The presence of EMFD in the poleward flank of the Hadley cell is accompanied by negative eddy heat flux convergence (EHFC) because the eddies are baroclinic in origin (colors, left column of Figure \ref{fig:HSheat}).  In the poleward flank of the Hadley circulation, heat diverges from low levels and converges at upper levels, indicating the slantwise, poleward heat transport of baroclinic eddies.  The magnitude of EHFC at the Hadley cell edge is less than 1 K day$^{-1}$ and is confined to low levels. Newtonian cooling (colors, right column of Figure~\ref{fig:HSheat}) is smaller in magnitude and broader in extent compared to EHFC, with net heating in the deep Tropics and cooling in the extratropics.

\begin{figure*}
\includegraphics[width=\textwidth]{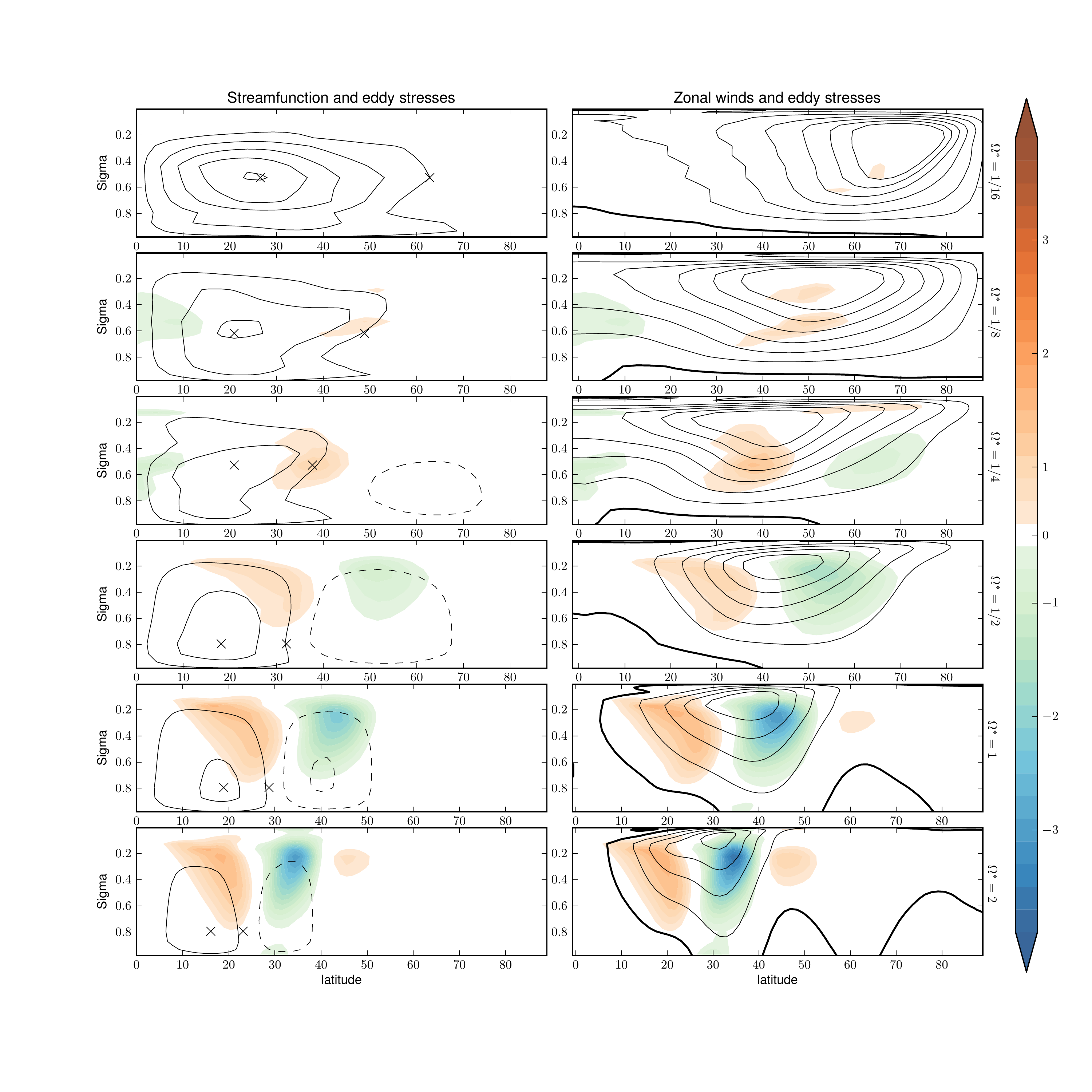}
\caption{As in Figure \ref{fig:HS}, but for the convective-relaxation model.
}
\label{fig:DryConv}
\end{figure*}

In the convective-relaxation model, the Hadley cells also generally widen and strengthen with decreasing rotation rate (black contours, left column of Figure \ref{fig:DryConv}).  The magnitudes of EMFD (colors in Figure \ref{fig:DryConv}) are comparable to those of the Held-Suarez-like simulations.  Perhaps less apparent and unlike the Held-Suarez model, EMFD strengths are non-monotonic in rotation rate (see Figure \ref{fig:caltech}(d)).  Zonal-mean zonal winds (black contours in right column) generally spread poleward and strengthen along with the Hadley cell.  The $\Omega^*=2$ case appears to have developed a multi-jet structure, as is evident in the zonal winds and EMFD.

\begin{figure*}
\includegraphics[width=\textwidth]{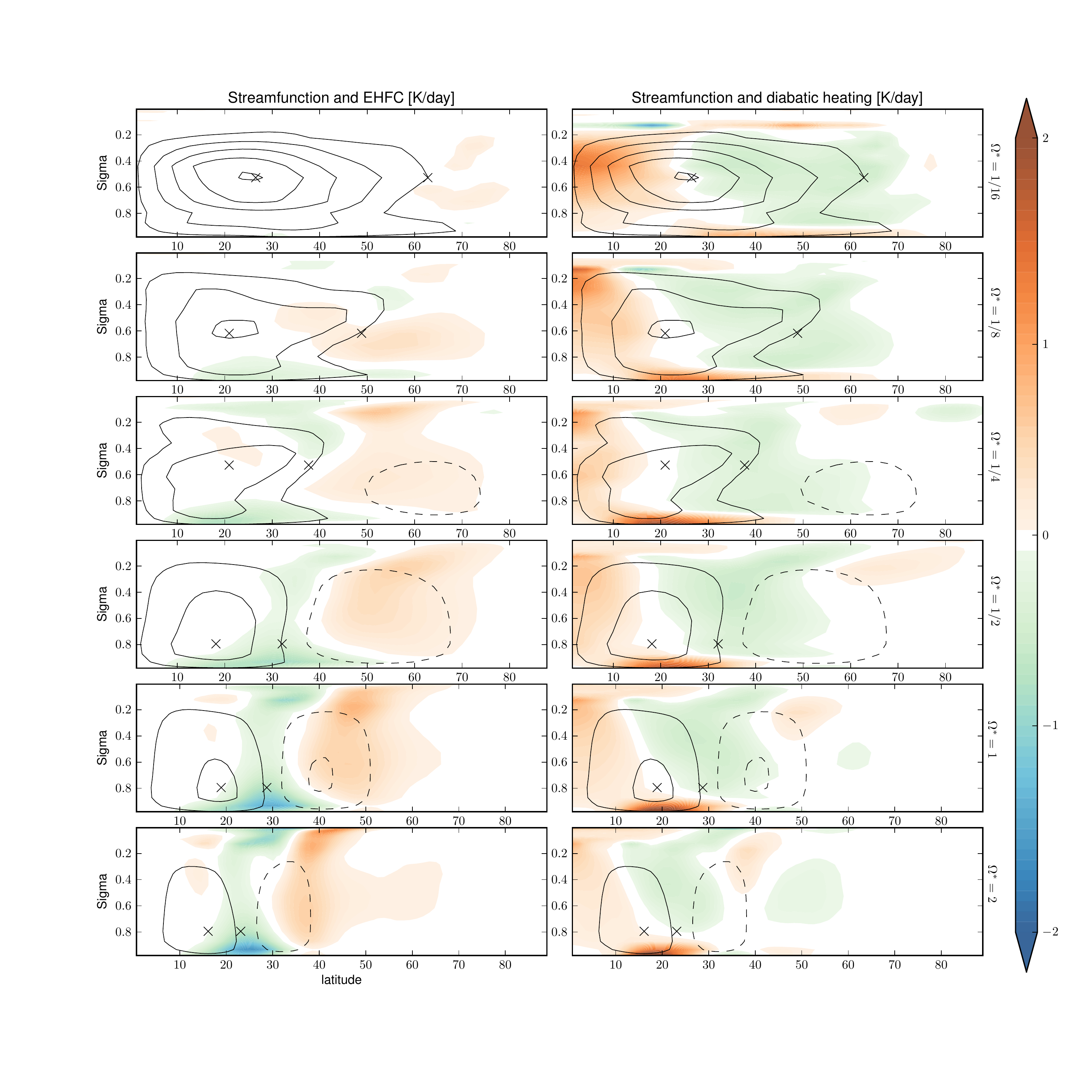}
\caption{As in Figure~\ref{fig:HSheat} but for the convective adjustment model.}
\label{fig:Caltechheat}
\end{figure*}

Eddies generally cool at the edge of the Hadley cell and heat in the extratropics (EHFC; colors in Figure \ref{fig:Caltechheat}), similar to the Held-Suarez-like simulations.  However, the slantwise heat transport is not as clear, especially for $\Omega^*\ge 1$.  Compared to the Held-Suarez-like simulations (Figure \ref{fig:HSheat}), the magnitude of eddy heating in the convective-relaxation model simulations is stronger for $\Omega^*\ge 1/4$ and weaker for $\Omega^*\le 1/8$, giving these simulations a more monotonic dependence on rotation rate.  This trend suggests that baroclinic instability weakens with decreasing rotation rate.

\subsection{Downdraft-averaged static stabilities, effective heating rates, and vertical velocities}
\begin{figure}[htbp]
  \includegraphics[height=0.25\textheight,keepaspectratio]{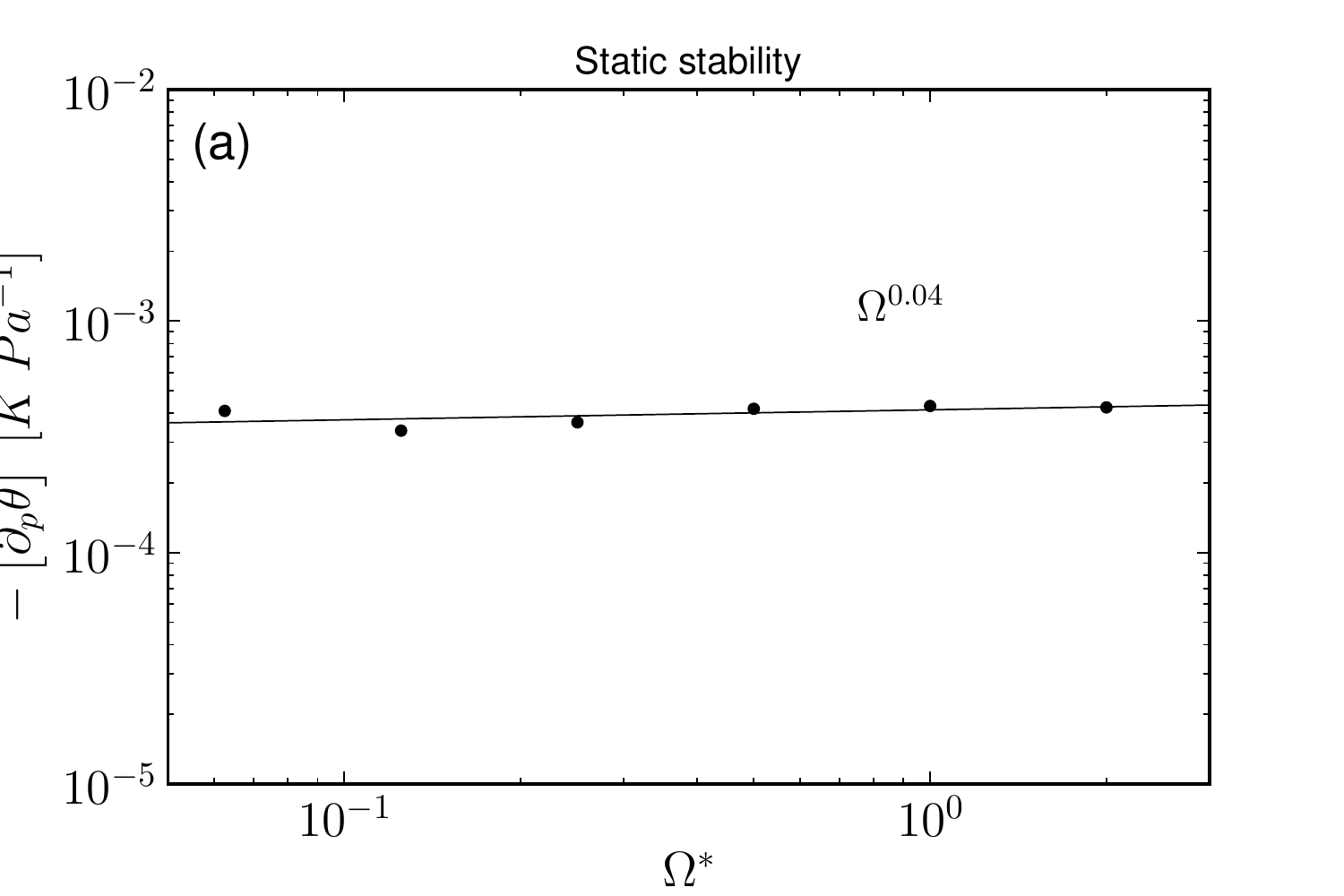}\\
  \includegraphics[height=0.25\textheight,keepaspectratio]{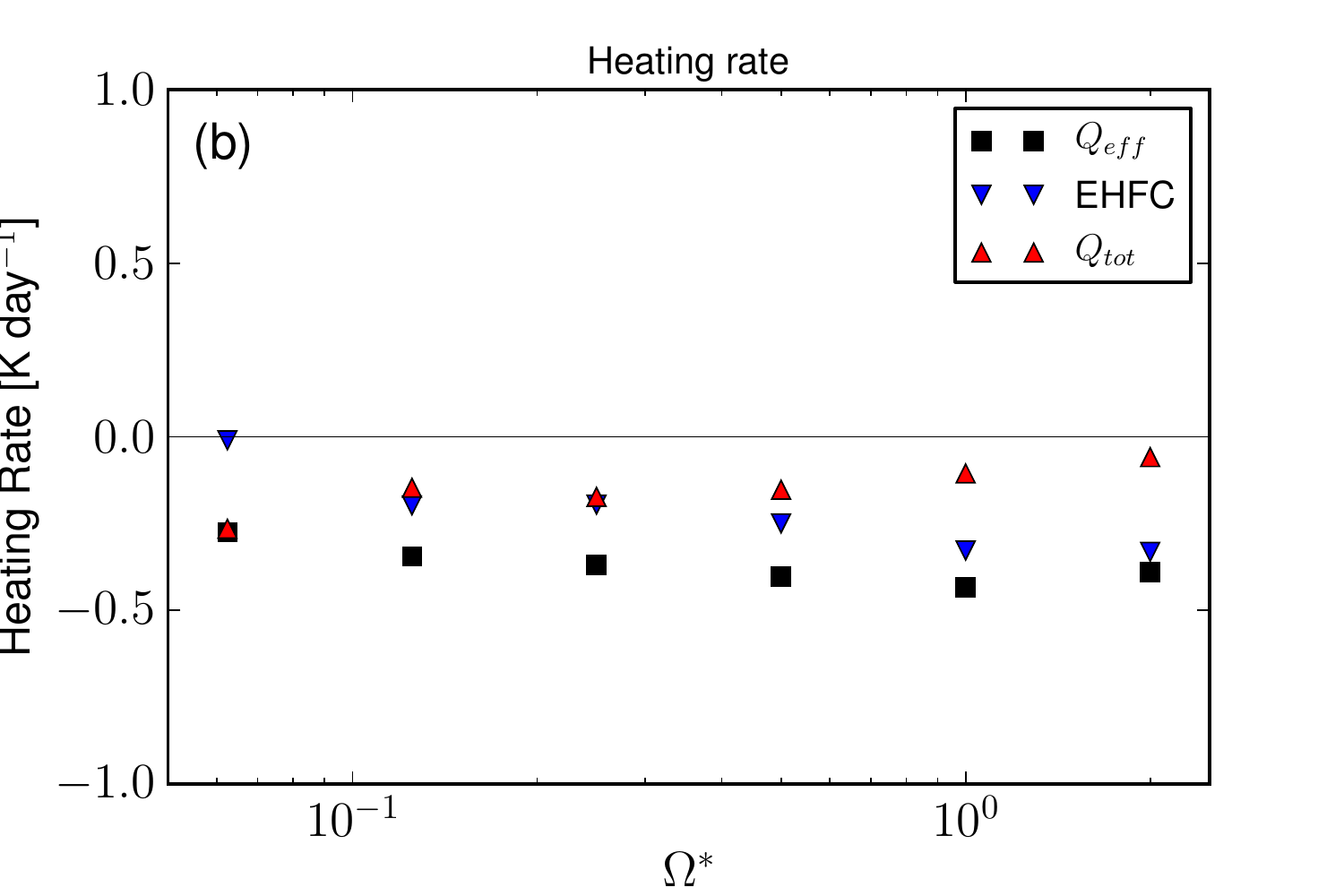}\\
  \includegraphics[height=0.25\textheight,keepaspectratio]{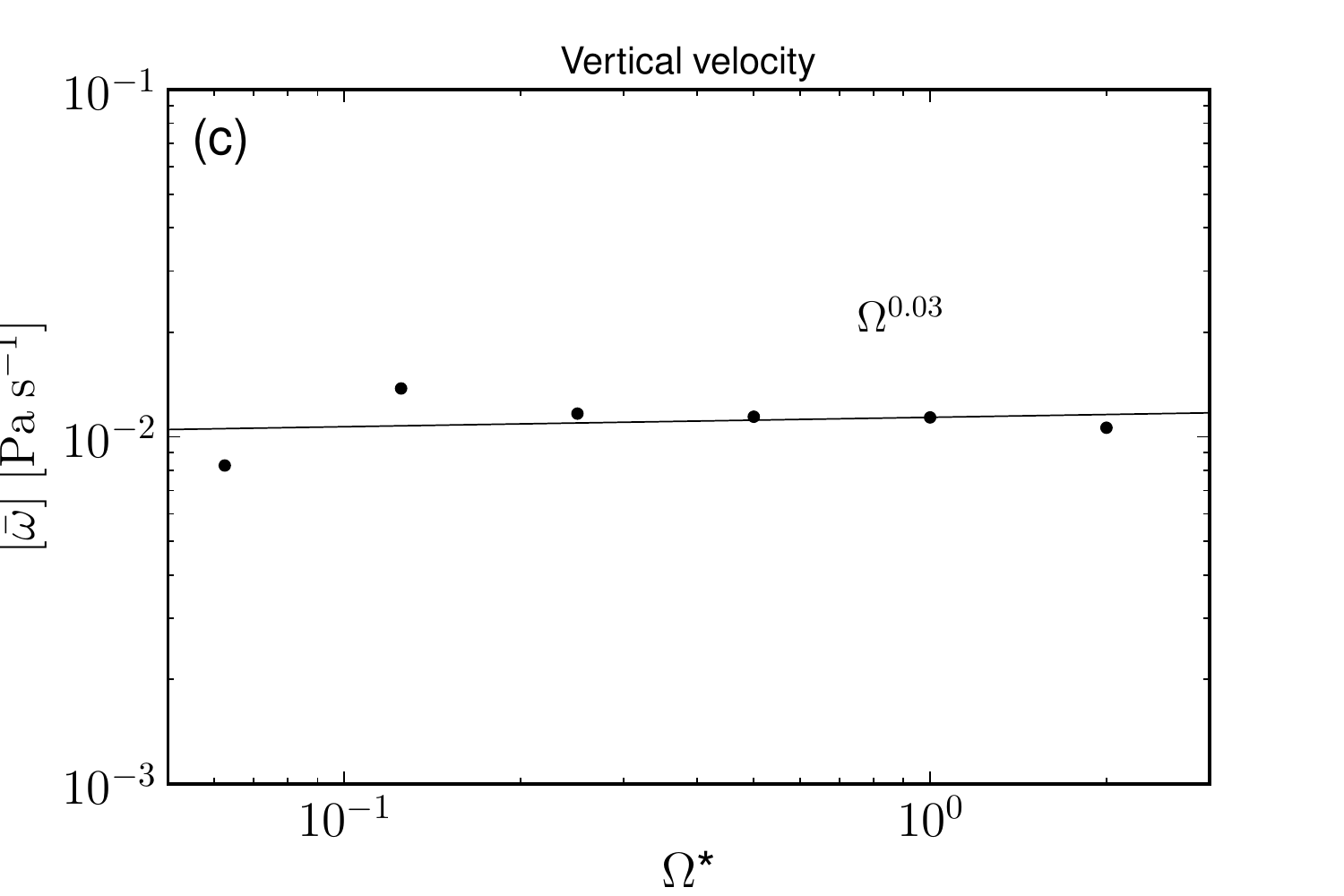}
\caption{(a) Zonal-mean static stability, $\partial_p\bar\theta$, at the level of the maximum streamfunction and averaged over the width of the downdraft for each of the Held-Suarez-like simulations.  (b) Heating rates for the Held-Suarez-like simulations at the level of the maximum Hadley cell mass flux and averaged over the downdraft.  Total heating rate ($Q_\textrm{eff}$, black squares); eddy heat flux convergence (EHFC, blue-downward triangles); and total diabatic heating (Qtot, red-upward triangles). (c) Zonal-mean vertical (pressure) velocities at the level of maximum Hadley cell mass flux and averaged over the downdraft.}
\label{fig:HSstabs}
\end{figure}

Across the Held-Suarez simulations, the static stability is nearly constant, and $Q_\mathrm{eff}$ depends very weakly on $\Omega$ (Figure~\ref{fig:HSstabs}).  EHFC and total diabatic heating, $Q_{\rm tot}$, have weak, but opposite dependence on rotation rate. Remarkably, these dependencies nearly offset one another to give a constant $Q_{\rm{eff}}$.  Thus the ratio of $Q_{\rm{eff}}$ to static stability in (\ref{eq:omegathermo}) is also nearly constant.

\begin{figure}[htbp]
  \includegraphics[height=0.25\textheight,keepaspectratio]{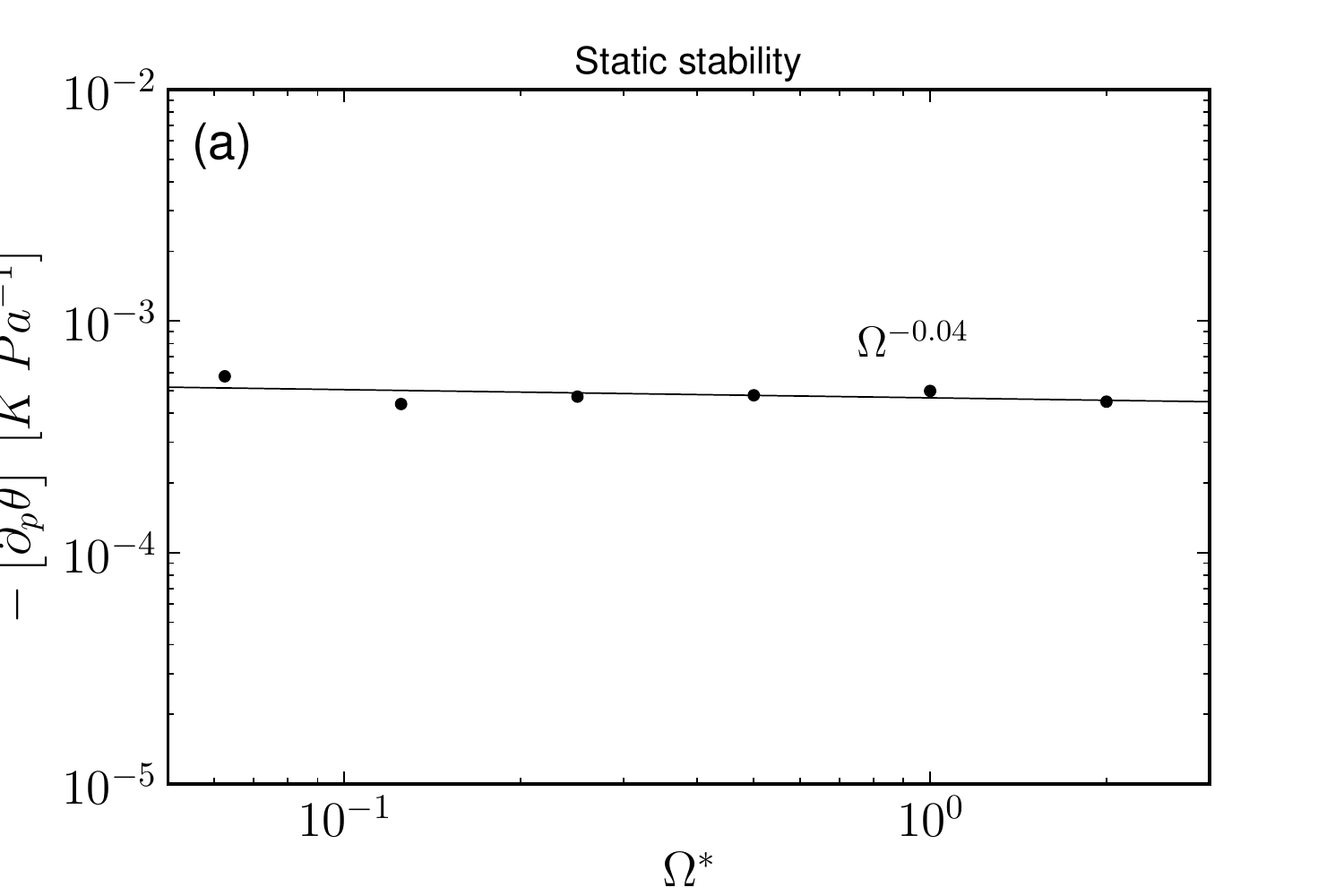}\\
  \includegraphics[height=0.25\textheight,keepaspectratio]{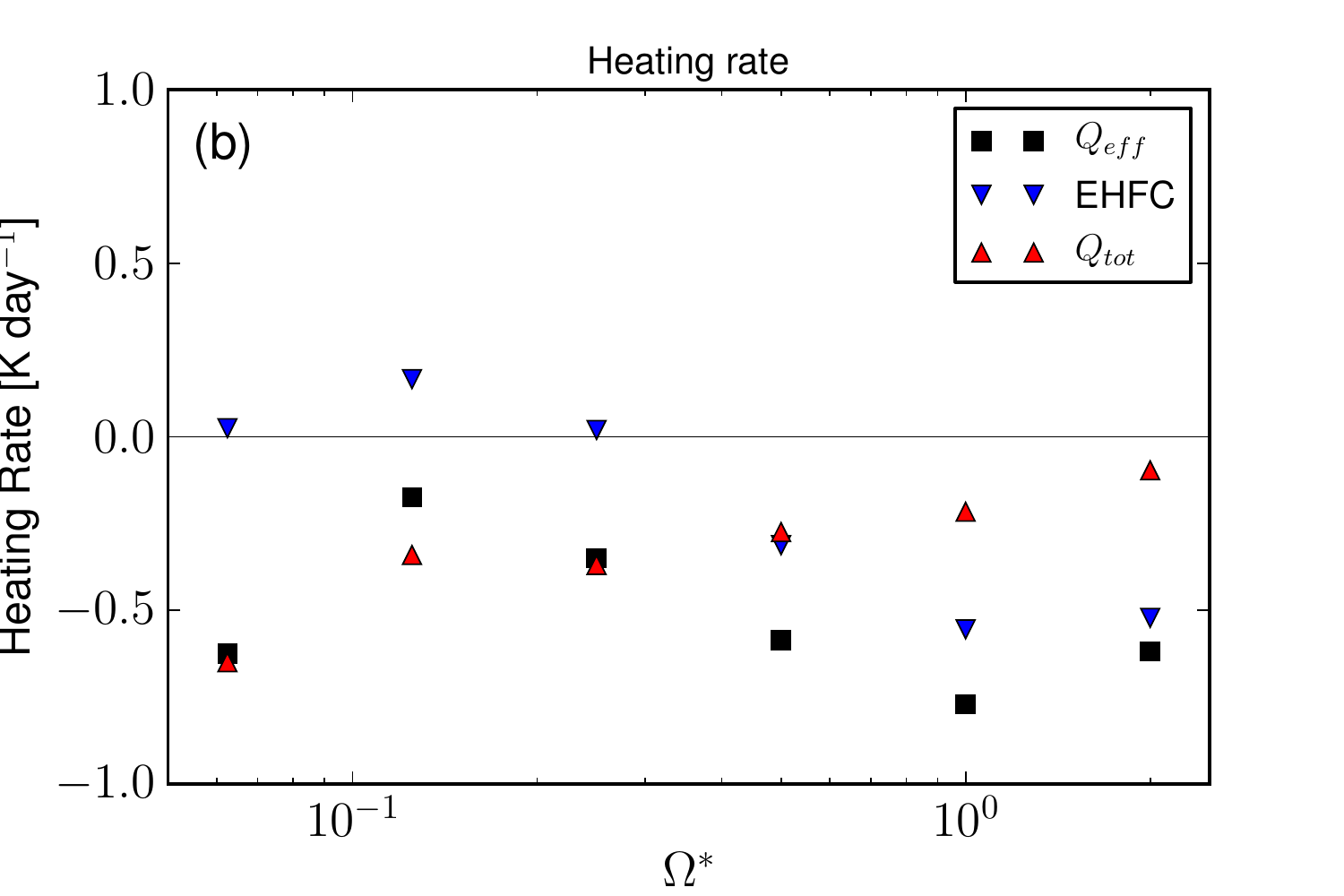}\\
  \includegraphics[height=0.25\textheight,keepaspectratio]{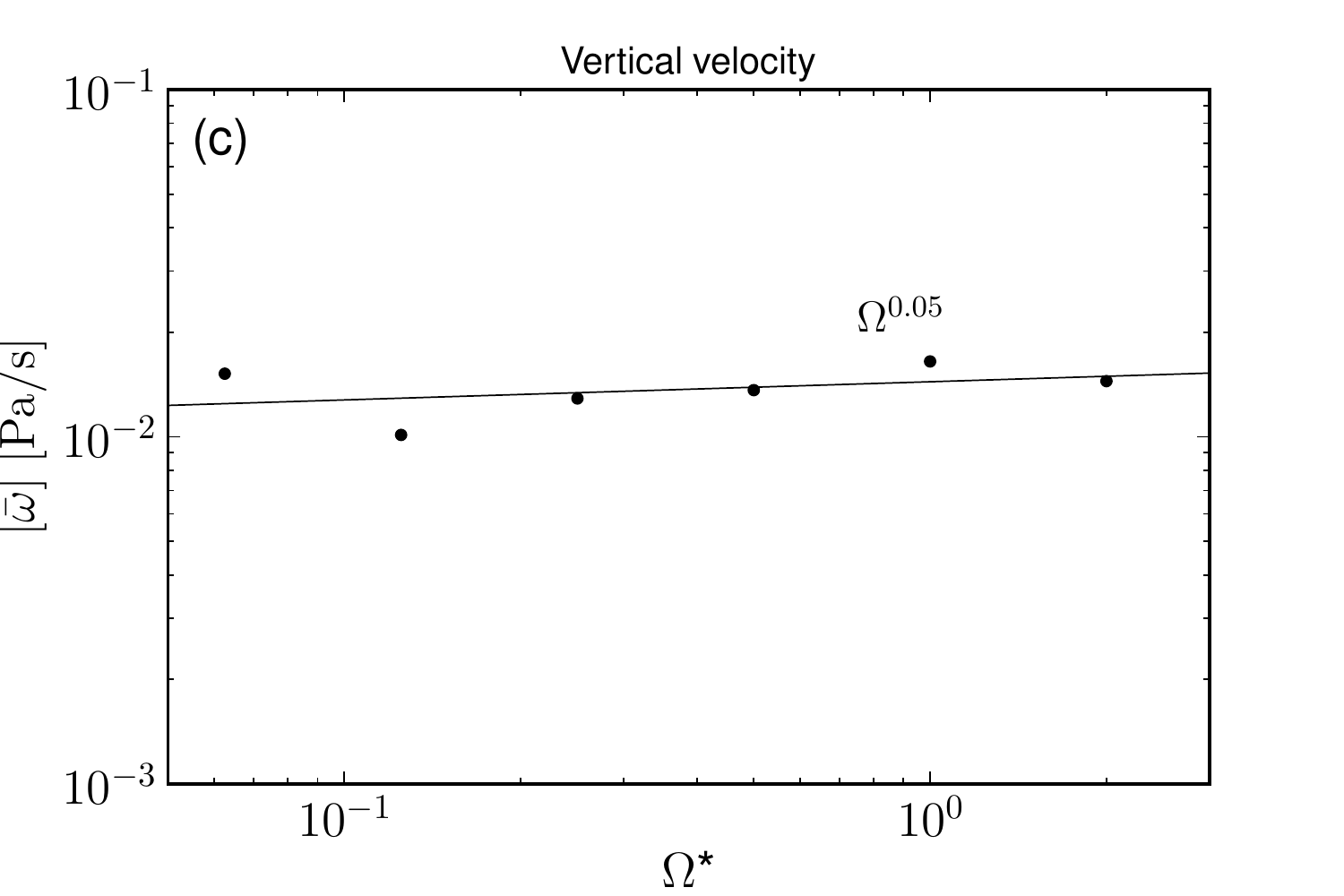}
\caption{As in Figure~\ref{fig:HSstabs}, but for the convective adjustment model.}
\label{fig:DCHR}
\end{figure}

Results are similar in the convective-relaxation model.  Downdraft-averaged static stability, effective heating, and vertical velocity are all very weakly dependent on $\Omega^*$ (Figure~\ref{fig:DCHR}).  Here again, compensation between $Q_{\rm{tot}}$ and EHFC is evident, but the reasons for this remain unclear.

\subsection{Hadley cell strength, downdraft width, eddy stresses, and bulk Rossby number results}
\begin{figure*}[ht]
\begin{center}
\includegraphics[width=25pc]{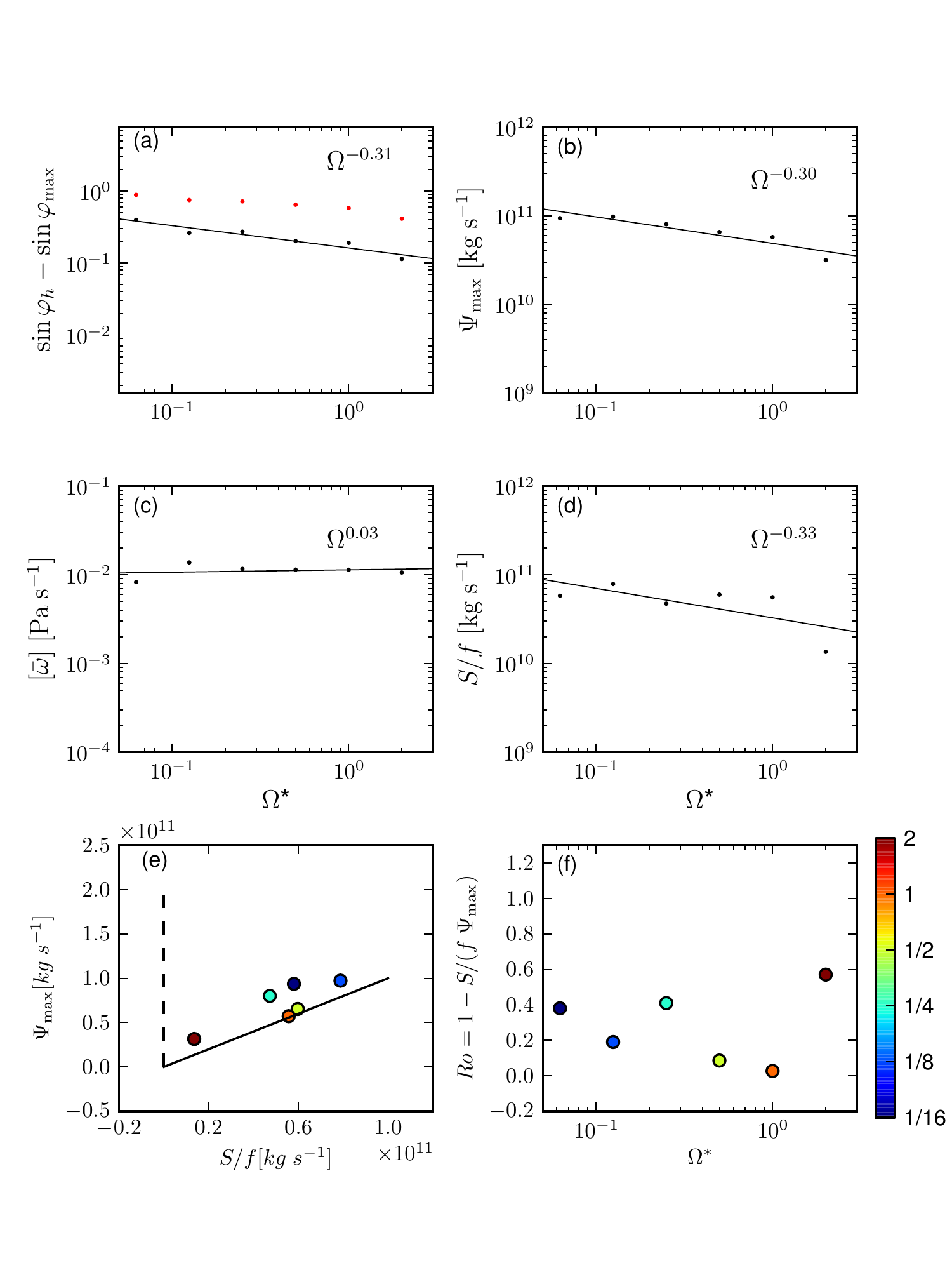}
\caption{Results from Held-Suarez-like forcing experiments with varying rotation rates (scaled to the Earth's value).  (a) The width of the downdraft of the Hadley circulation (black dots), best-fit power-law (black line), and total width of the Hadley cell (red dots).  (b) The meridional-mean mass flux over the downdraft at the level of the maximum of the Hadley circulation [kg s$^{-1}$] (black dots) and best-fit power-law.  (c) Zonal- and meridional-mean vertical pressure velocities [Pa s$^{-1}$] averaged over the width of the Hadley cell downdraft (black dots) and best-fit power-law (black line).  (d) Bulk eddy stresses divided by the coriolis parameter at the maximum of the Hadley circulation, as defined in (\ref{eq:psimax}) (black dots), and best-fit power law (black line).  (e) The quantity $S/f$ averaged at the maximum of the Hadley circulation plotted against the maximum mass flux of the Hadley cell, $\Psi_{\rm max}$.  The solid line indicates the values that would correspond to $Ro=0$, and the dashed line would correspond to $Ro=1$. (f) The value of the bulk Rossby number, $Ro = 1-S/(f \ \Psi_{\rm max})$, as a function of $\Omega^*$.}
\label{fig:HSlike}
\end{center}
\end{figure*}

Figure \ref{fig:HSlike}(a)-(d) displays the scaling of key characteristics of the Hadley cells with planetary rotation rate for the Held-Suarez model simulations, including the width of the downdraft, the strength measured by the maximum mass flux, the vertical pressure velocities averaged over the downdraft, and the bulk eddy stresses divided by the coriolis parameter, $S/f$, at the location of the maximum Hadley cell mass flux.  As noted earlier and in other studies \cite[e.g.,][]{Williams_88a,Williams_88b,Walker_Schneider06,DiasPinto_Mitchell14}, the Hadley cell becomes wider and stronger as the rotation rate decreases.  As described in the previous section, the downdraft velocity is nearly constant, albeit with some scatter, for the entire range of simulated rotation rates (Figure \ref{fig:HSlike}(c)).
To summarize, this is consistent with the thermodynamic constraint implied by (\ref{eq:omegathermo}) and Figure~\ref{fig:HSstabs}, namely that for constant $Q_{\rm{eff}}$ and static stability, vertical velocities must also be constant (the ``omega governor'').   Figure \ref{fig:HSlike}(d) shows the scaling of the bulk eddy stresses divided by the coriolis parameter, $S/f$, at the location of the maximum Hadley circulation mass flux, which roughly follow a power-law $\sim\Omega^{-0.33}$.  In our theory, the Hadley cell mass flux must be consistent with $S/f$.  If vertical velocities are fixed as Figure \ref{fig:HSlike}(c) demonstrates, then according to (\ref{eq:psimax}) and (\ref{eq:2psi}), both the strength of the Hadley cell, $\Psi_{\rm max}$, and the width of the downdraft, $\sin\varphi_h-\sin\varphi_m$, should have the same power-law scaling, inherited from the eddy stresses, $S/f$.  Figures \ref{fig:HSlike}(a), (b) and (d) show that the downdraft width ($\sim\Omega^{*-0.31}$), Hadley cell mass flux ($\sim\Omega^{*-0.30}$) and eddy stresses ($\Omega^{*-0.33}$) have similar power-law scalings.  This is consistent with the low-$Ro$ theory (\ref{eq:2psi}): eddy stresses create momentum demand in the downdraft region, and with fixed Hadley cell downdraft velocity, the cell must respond proportionally by widening or narrowing to increase or decrease mass flux based on the eddy stress demand.

To visualize how the eddy stresses contribute to the mass flux,  Figure \ref{fig:HSlike}(e) shows the Hadley cell strength, $\Psi_{\rm max}$, plotted against the eddy stresses, $S/f$, following \cite{Singh_Kuang16}.  From (\ref{eq:psimax}), these quantities are proportional to one another with the factor $1-Ro$ multiplying the Hadley cell mass flux.  If $Ro=0$, the bulk eddy stresses are equal to the Hadley cell mass flux (the solid line of Figure \ref{fig:HSlike}(e)).  If $Ro=1$, as would be the case in the axisymmetric, nearly-inviscid limit, the Hadley cell mass flux is independent of the eddy stresses (the dashed line of Figure \ref{fig:HSlike}(e)).  The values for our suite of simulations are plotted in Figure \ref{fig:HSlike}(e), and color-coded by their values of $\Omega^*$.  All of our simulations stay closer to the $Ro=0$ line than the $Ro=1$ line, regardless of the value of rotation rate.  The bulk Rossby number as diagnosed from (\ref{eq:psimax}), $Ro=1-S/(f \ \Psi_{\rm max})$, is shown as a function of $\Omega^*$ in Figure \ref{fig:HSlike}(f).  There is considerable scatter in $Ro$ values, however all but the fastest rotation rate, $\Omega^*=2$, have $Ro\le0.5$.  The reason for the scatter in $Ro$ is not obvious, however the sequence of mass streamfunctions and eddy momentum flux divergences in Figure \ref{fig:HS} suggest circulation changes -- for instance a stacked, double-maximum structure in the streamfunction -- corresponding to the jumps in $Ro$.  Indeed, the stacked cells appear to be split at or near the top of the boundary layer, $700$hPa, possibly suggesting an important role for boundary layer processes.

\begin{figure*}
\begin{center}
\includegraphics[width=33pc]{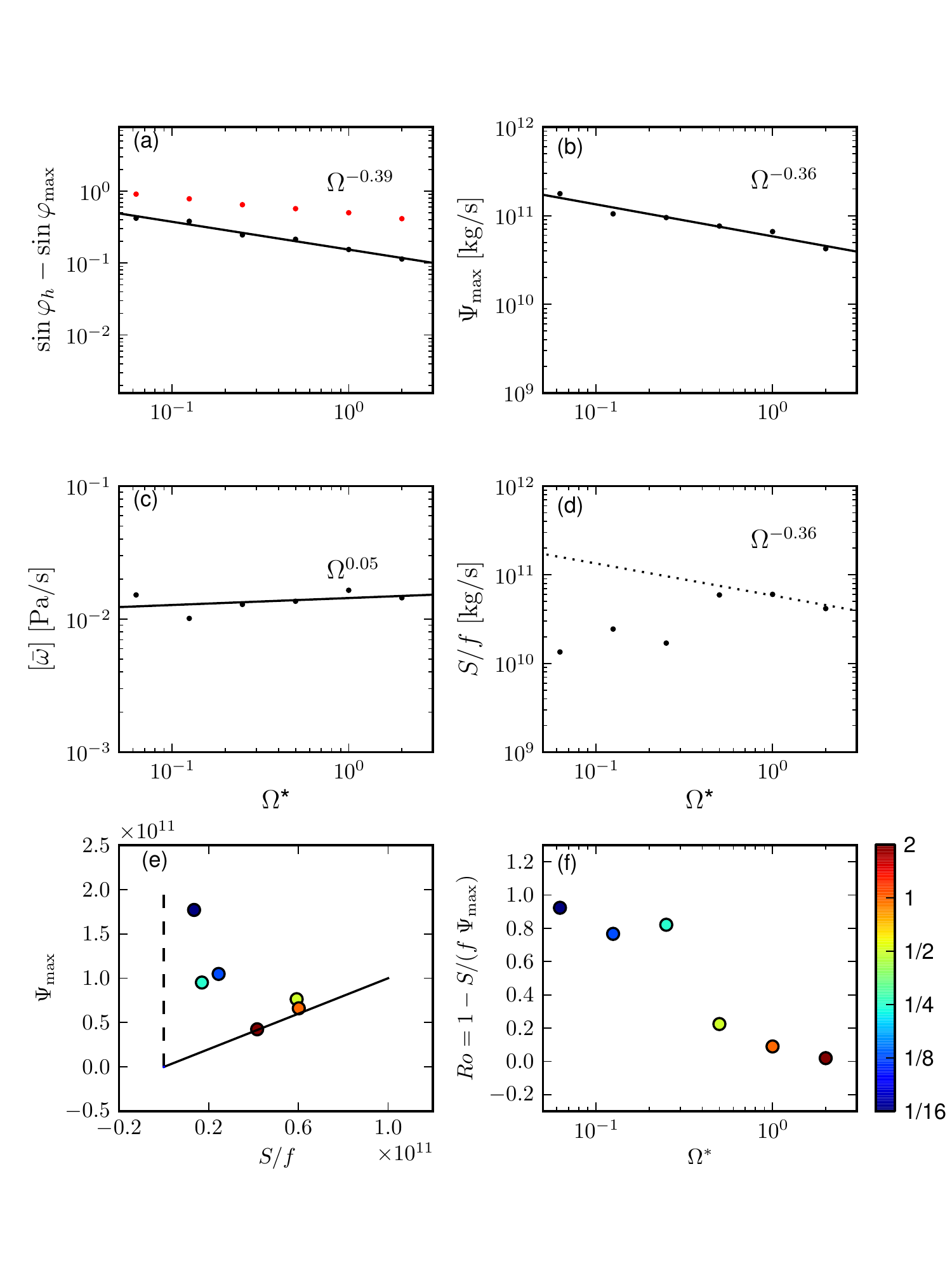}
\caption{Same as in Figure \ref{fig:HSlike}, but for the convective-relaxation model.  The dotted line in panel (d) is the same as the solid line in panel (b).}
\label{fig:caltech}
\end{center}
\end{figure*}

For the convective-relaxation simulations, Figure \ref{fig:caltech}(c) shows that the vertical velocities averaged over the downdraft are also roughly independent of rotation rate.  Panels (a) and (b) show the width and strength increasing with decreasing rotation rate, and with similar power-laws of $\sim\Omega^{*-0.39}$ and $\sim\Omega^{*-0.36}$, respectively, similar to those of the previous model. But unlike the Held-Suarez case the widths and strengths are not straightforward imprints of the $S/f$ scaling (panel d).  Instead, there is a hint of the same scaling for $S/f$ as the Hadley cell strength (dotted line in panel d) for the fastest three rotation rates, then a large deviation to smaller values of $S/f$ for smaller rotation rates.  As planetary rotation slows, the circulation strength deviates from $S/f$ and becomes more angular-momentum-conserving (i.e., $Ro$ increases).

In Figure \ref{fig:caltech}(e) and (f), we see that the three largest rotation rates are near the $Ro=0$ line, in the range that we expect our theory to be valid.  The smaller values of rotation, however, have $Ro\sim1$, and these are the same cases that deviate from the power law in $\Omega^{*}$ expected for $Ro\ll 1$ (Figure~\ref{fig:caltech}(d)).  Remarkably and for reasons still unclear (although see Section \ref{sec:disc}), the increase in $Ro$ exactly offsets the decrease in $S/f$ so that the power laws of the widths and strengths remain the same across low and high values of $\Omega^*$ (Figure \ref{fig:caltech}(a) and (b)).

This case, Figure \ref{fig:caltech}, illustrates an important feature of the omega governor, namely that its validity doesn't depend on the value of $Ro$.  The definition of the streamfunction, (\ref{eq:psimaxomega}), makes clear that if the omega governor applies, the strength, $\Psi_{\rm max}$, and the width of the downdraft, $\sin\varphi_{\rm max}-\sin\varphi_h$, will have the same scaling with rotation rate.  It then remains to determine the scaling of the downdraft width or strength by some other constraint.  Additionally, the existence of a robust, $Ro$-independent power-law scaling suggests there may exist a yet-to-be-determined driving mechanism other than eddy stresses or a thermally direct circulation, on which we provide some speculative arguments further in Section \ref{sec:disc}.

\section{Simulations of an idealized moist atmosphere}
\label{sec:moist}
Considering moist atmospheres, WTG constraints on static stability certainly continue to hold.  But with a latent heating distribution that depends interactively on the circulation itself due to evaporation and condensation of moisture, the prospects for a fixed $Q_\mathrm{eff}$ become even more remote.  And the arguments of \cite{jansen_equilibration_2013} regarding the extratropical static stability also presume a dry atmosphere.  Nevertheless, in an attempt to bridge our results from dry atmospheres to Earth's moist one, we now explore a final suite of simulations performed with an idealized aquaplanet model.

The aquaplanet features idealized, gray radiation and simplified moist physics and is also based on the same FMS spectral dynamical core \citep{Frierson_etal06}.  Briefly, the model uses a fixed, constant longwave optical depth\footnote{Note the original formulation of longwave optical depth had latitude-dependence, which we remove.} to solve the radiative transfer equation, a simplified convective relaxation scheme for handling moist physics, and a simplified Monin-Obukhov turbulence parameterization.  The surface is assumed to be a slab of uniform thickness and unlimited water supply.  Simulations are run at T42 resolution with 25 vertical levels, and averages are taken over the final 360 of 1440 days.

\begin{figure*}
\includegraphics[width=\textwidth]{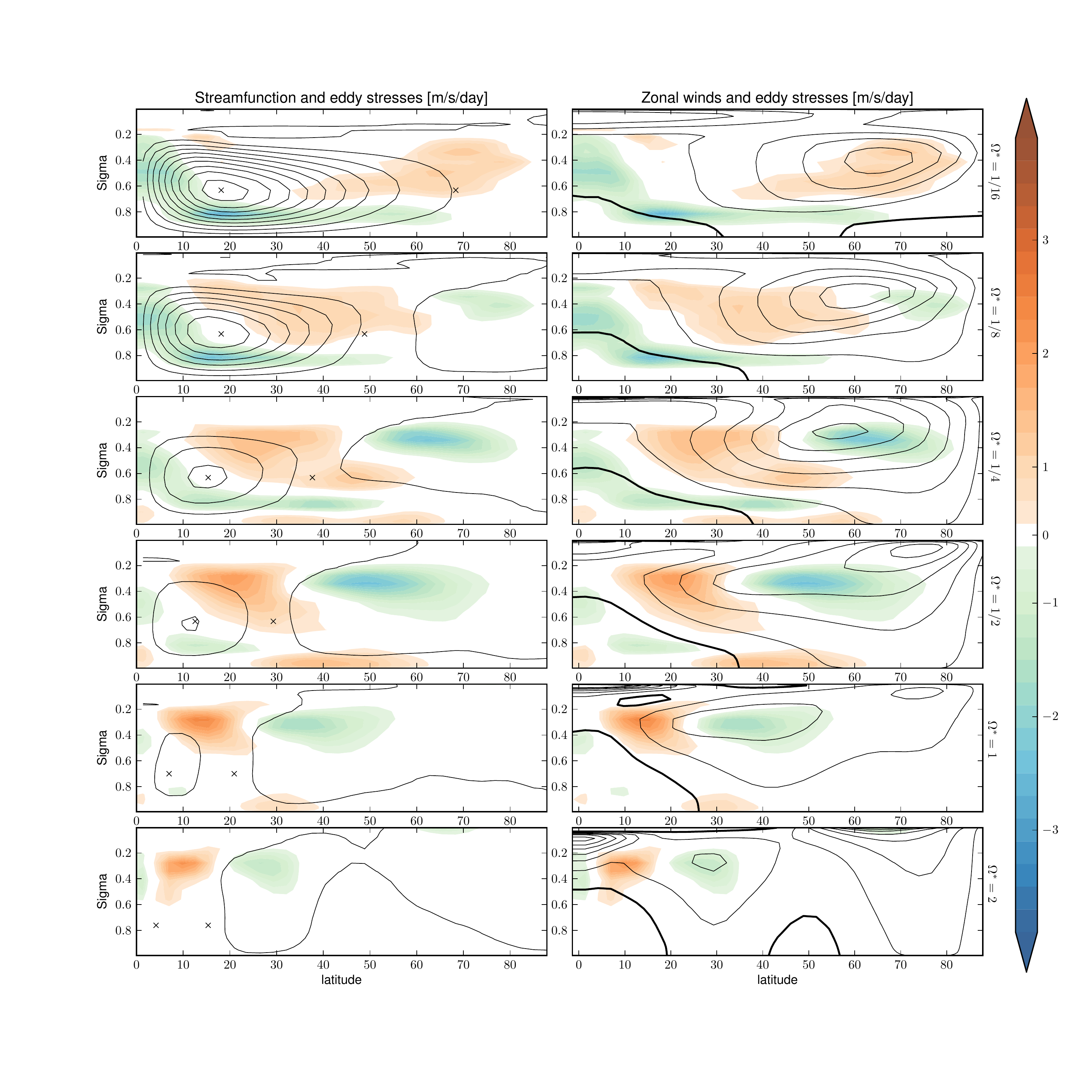}
\caption{As in Figure \ref{fig:HS}, but for the idealized moist GCM.  
}
\label{fig:aquaplanet}
\end{figure*}

\begin{figure*}
\includegraphics[width=\textwidth]{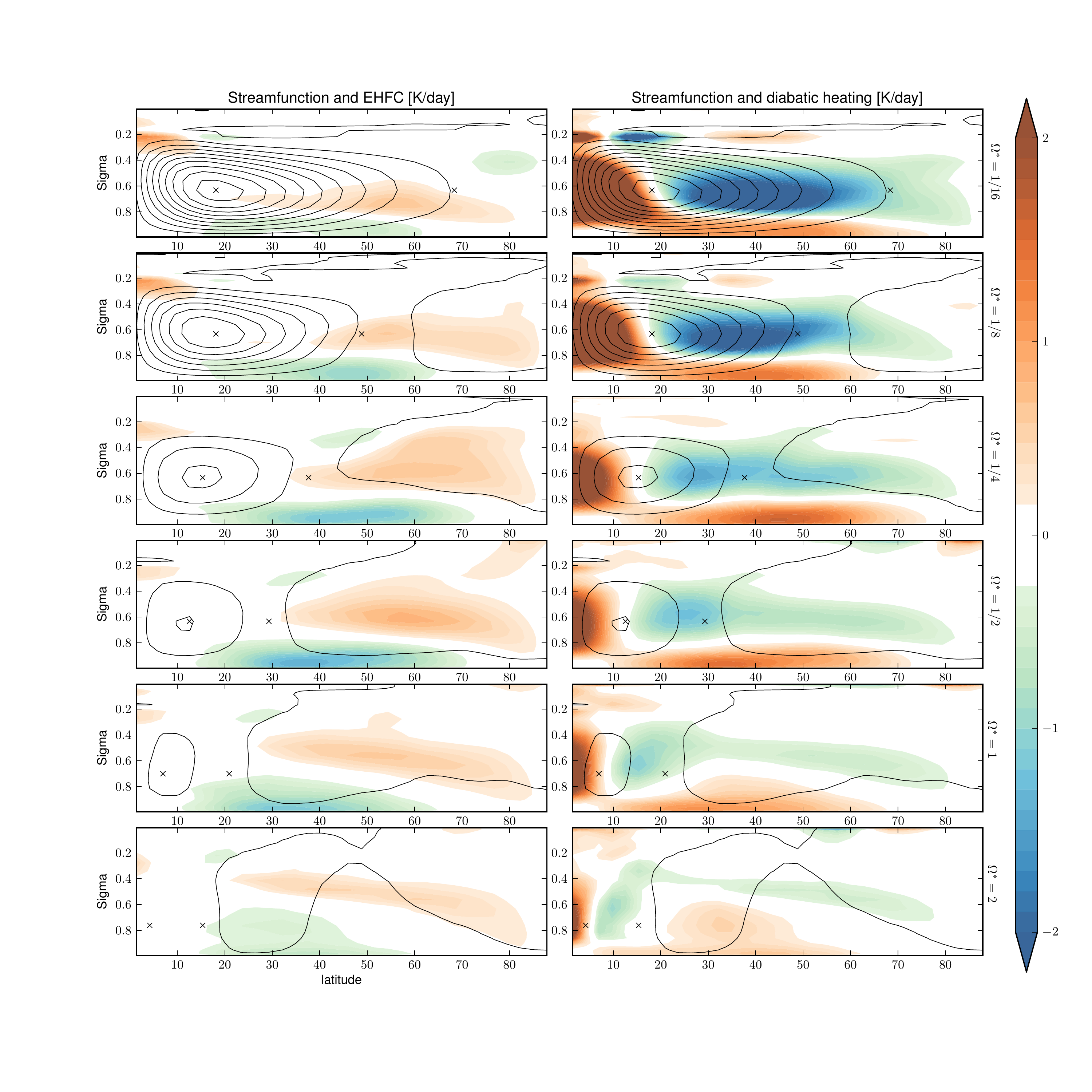}
\caption{As in Figure \ref{fig:HSheat} but for the idealized moist GCM.
}
\label{fig:aquaheat}
\end{figure*}

\begin{figure}
  \includegraphics[height=0.25\textheight,keepaspectratio]{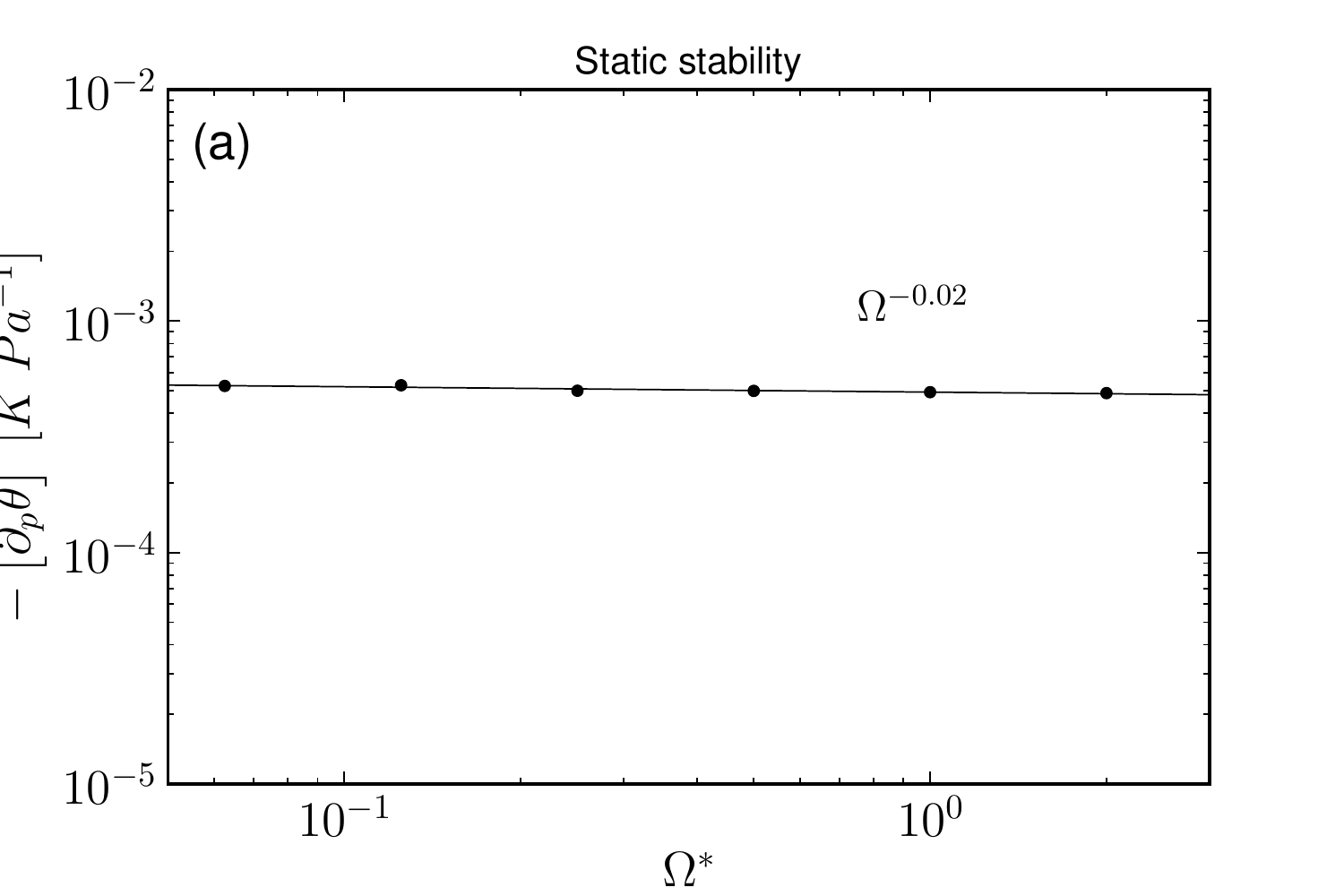}\\
  \includegraphics[height=0.25\textheight,keepaspectratio]{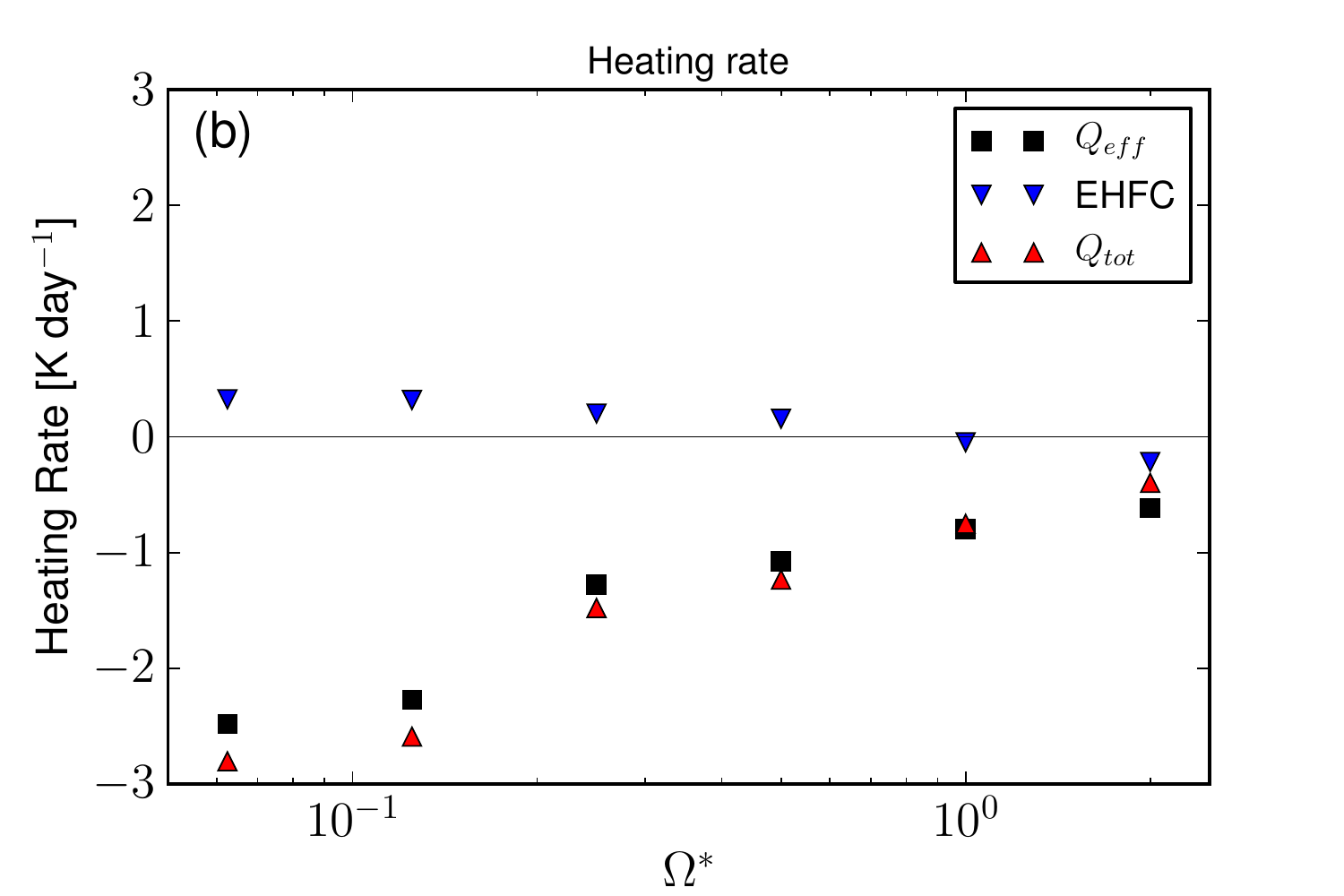}\\
  \includegraphics[height=0.25\textheight,keepaspectratio]{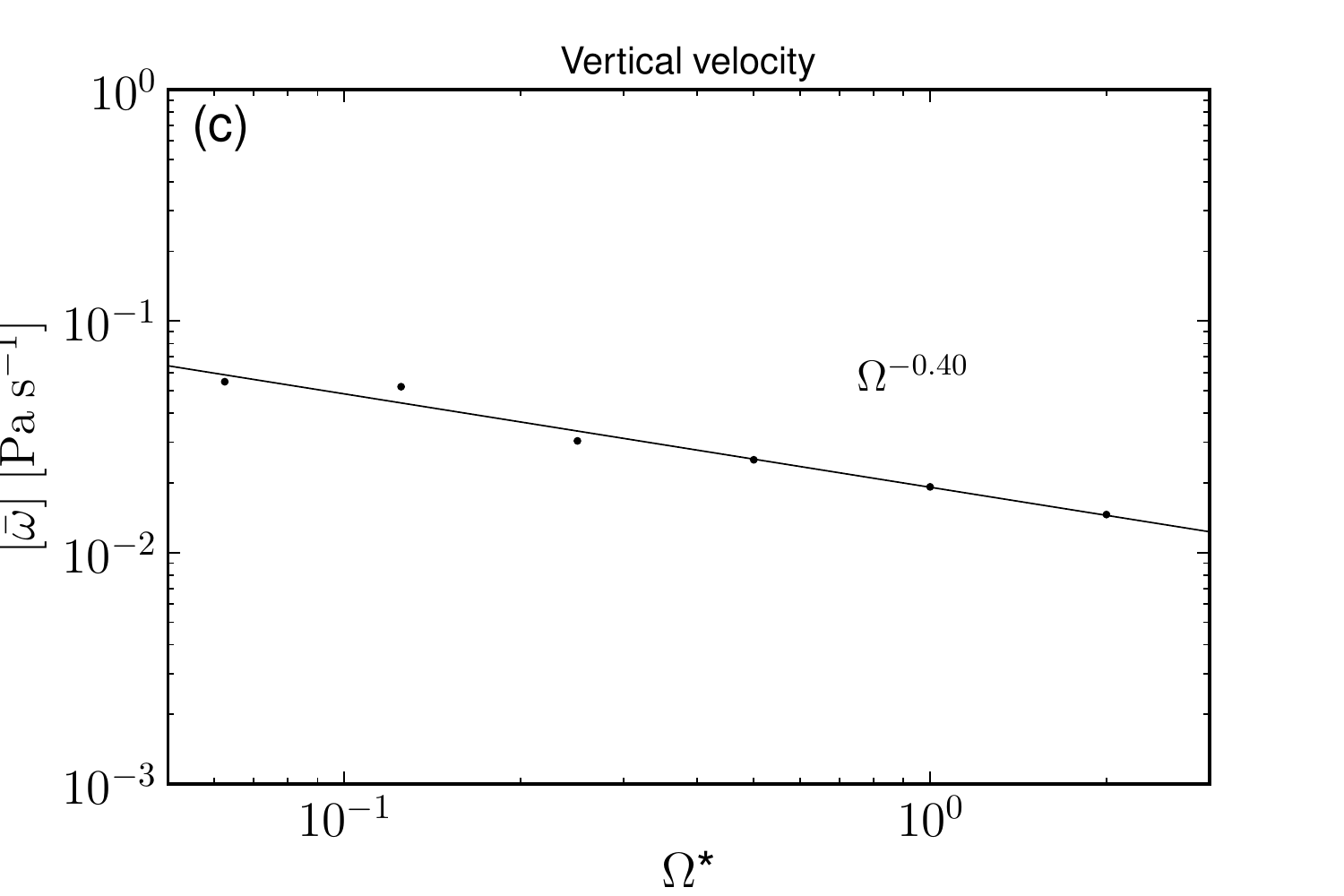}
\caption{As in Figure \ref{fig:HSstabs} but for the idealized moist GCM.}
\label{fig:aquastabs}
\end{figure}

\begin{figure*}
\begin{center}
\includegraphics[width=33pc]{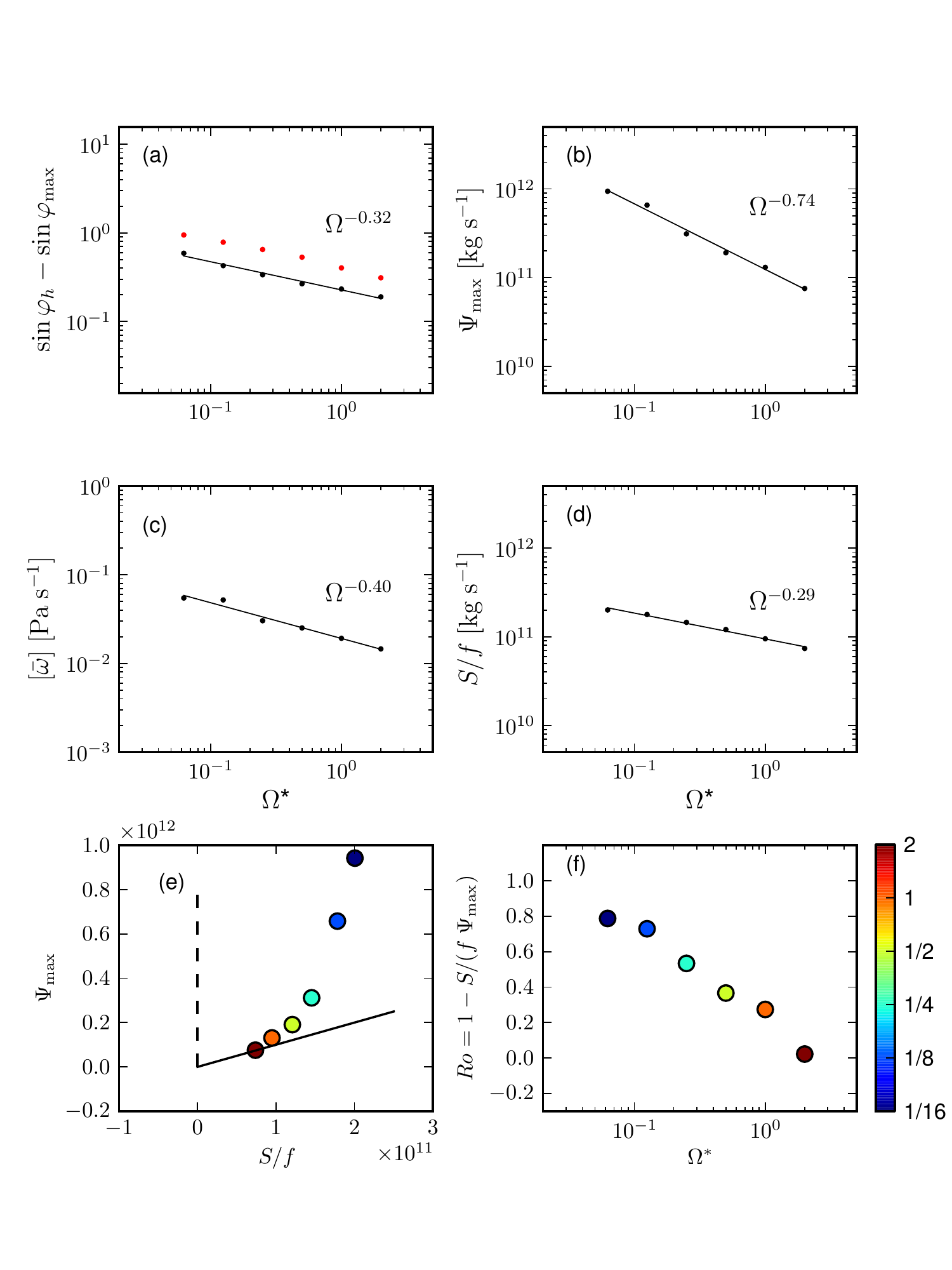}
\caption{As in Figure \ref{fig:HSlike}, but for the idealized moist GCM.}
\label{fig:aqua}
\end{center}
\end{figure*}

The dependence of the aquaplanet general circulation on rotation rate is qualitatively consistent with that of the other two model forcings (Figure \ref{fig:aquaplanet}).  The Hadley cell widens and strengthens with decreasing rotation rate, zonal-mean zonal winds increase in magnitude and spread poleward, and EMFD generally follows with the edge of the Hadley cell.  The magnitude of EMFD is, however, considerably smaller than for the other model forcings, and the Hadley cell is stronger than the other forcings at $\Omega^*=1$ and 2.  There is also a hint of a double-jet structure in these cases, as well as strong, upper-level superrotation.

The magnitudes of EHFC and diabatic heating (colors in Figure \ref{fig:aquaheat}) are generally larger than in the Held-Suarez-like and dry-convective simulations.  And compared to the dry models, the aquaplanet EHFC does not substantially weaken with decreasing rotation rate, although it does move poleward with the expanding Hadley cell.  Qualitatively, there appears to be slantwise heat transport as in the Held-Suarez-like case.

Turning to thermodynamic considerations in the Hadley cell downdraft, static stabilities are, as in the dry models, independent of $\Omega^*$ (top panel of Figure \ref{fig:aquastabs}).  The effective heating, on the other hand, is \emph{not} fixed, rather it scales inversely with $\Omega^*$ (middle panel of Figure \ref{fig:aquastabs}).  Diabatic heating, $Q_{\mathrm{tot}}$, dominates over EHFC for all $\Omega^*$, which simplifies the interpretation of $Q_\mathrm{eff}$.  Note that EHFC only accounts for dry eddy effects, and we include all moist effects in $Q_\mathrm{eff}$, which also includes radiative heating by gray radiation.  Thus the dependence of $Q_\mathrm{eff}$ on $\Omega^*$ in the aquaplanet simulations could be due to changes in moisture flux divergence, convection and/or gray radiative transfer. Downdraft-averaged vertical velocities inherit the $\Omega^*$-dependence of the effective heating, as they must (Figure \ref{fig:aquastabs}(c)).  Thus, moisture muddles our theoretical interpretation by introducing $\Omega^*$-dependence to $Q_{\mathrm{eff}}$, which renders the fixed-$\left[\bar\omega\right]$ assumption of the omega governor invalid.

Because the downdraft velocity is no longer independent of rotation rate (Figure \ref{fig:aquastabs}(c)), the magnitude of the convective heating in general depends on the local circulation, which is an effect we have not accounted for (this would be relevant for the convective-relaxation model as well). It may also be that the addition of dynamical latent energy fluxes to the aquaplanet forcing cause the rotation-rate dependence.  Regardless of the exact cause, it seems the presence of moisture may limit the validity of the omega governor.

The value of $Ro$ steeply descends with rotation rate (Figure \ref{fig:aqua}(f)), so we might expect complicated dependence of the Hadley cell strength and downdraft width on $\Omega^*$.  The eddy stresses scale as $S/f \sim\Omega^{*-0.29}$, and the width ($\sim\Omega^{*-0.32}$; Figure \ref{fig:caltech}(a)) shares nearly the same scaling.  Note importantly that aquaplanet downdraft widths scale almost identically to the two dry models (compare Figures \ref{fig:HSlike}(a), \ref{fig:caltech}(a) and \ref{fig:aqua}(a)).  But in a major departure from the dry models, this does not match the scaling of the Hadley cell strength ($\sim\Omega^{*-0.74}$; Figure \ref{fig:caltech}(b)).  Recall, however, that our theory requires $Ro\ll1$ and $\left[\bar\omega\right]$ must be independent of $\Omega^*$, and neither condition is met for this model.  Because of this combination of factors, we do not expect our theory to account for the model behavior.  However, it can be seen that the downdraft velocity times the width has approximately the same scaling as $\Psi_{\rm max}$, as is always required by (\ref{eq:psimaxomega}).  Understanding the scaling of eddy stresses, $S/f$, relative to $\Psi_{\rm max}$ requires knowledge of the scaling of $(1-Ro)$, and this is left to future work.

The aquaplanet model simulations are outliers; our small-$Ro$ scaling (\ref{eq:2psi}) does not appear to hold even when the simulated $Ro$ is small, and $[\bar\omega]$ is not constant.  Even so, accounting for the dependence of $Ro$ on rotation rate makes the aquaplanet solutions consistent with the more general expression of the theory when $Ro$ is not small, (\ref{eq:full2psi}).  This is of course how it had to be, provided the other conditions leading to (\ref{eq:psimax}) are met, but it's nonetheless instructive to see these scalings work well even if vertical velocities are not constant.

\section{Discussion and conclusions}
\label{sec:disc}

\subsection{Results for the updraft}
The downdraft width does not directly constrain the total Hadley cell width.  What controls the updraft width remains actively researched \citep{Byrne_etal18}.  But a theory for downdraft width combined with a theory for the updraft width (or their ratio) would constitute a complete theory for the total cell width.

In principle, the same arguments leading to an omega governor in the downdraft are applicable in the updraft, since the same leading-order balance should apply.  In other words, as long as a three-term balance between vertical advection, diabatic heating, and eddy heat flux divergence holds, and so long as the diabatic plus eddy term divided by the static stability is fixed, the updraft vertical velocity will be fixed also.  However, taking the convective-relaxation model as an example, though the updraft-averaged static stability remains fixed across rotation rates in our simulations, the updraft-averaged $\omega$ varies strongly with rotation.  Lapse rates in the updraft of the Held-Suarez-like simulations are similarly invariant, but $Q_{eff}$ and vertical velocities are not. In contrast, aquaplanet simulations have essentially identical scalings in the updraft and downdraft (not shown). We leave the explanation of these phenomena to future work.

Nevertheless, as the overlain red dots in Figures \ref{fig:HSlike}(a), \ref{fig:caltech}(a), and \ref{fig:aqua}(a) showing the total Hadley cell width demonstrate, to first order in all three models the total cell width and downdraft width do share a similar scaling, even when moisture is present. As rotation rate decreases and the overall cell expanse grows, equivalent downdraft and total width scalings amount to either the updraft widening at about the same rate as the downdraft, or a fixed updraft width and a widening downdraft.

Other lines of argument link the updraft and downdraft widths, notably those of \cite{Watt-Meyer_Frierson19}: the narrower the updraft, the larger the planetary angular momentum values imparted to the free troposphere, therefore the larger the meridional shear as parcels move poleward (absent compensating changes in eddy stresses), and thus a more equatorward onset of baroclinic instability and with it the cell edge \citep{Held00,Kang_Lu12}.

\subsection{Role of Ekman pumping}
As noted earlier, all three models have nearly identical scalings for Hadley cell widths with rotation rate, $\sin\varphi_h\sim\Omega^{*-1/3}$ (Figures \ref{fig:HSlike}(a), \ref{fig:caltech}(a) and \ref{fig:aqua}(a)).  What is the origin of this universal scaling? And in the convective relaxation model, how does the strength decouple from eddy stresses in such a coordinated fashion (Figure \ref{fig:caltech}(b) and (d))?  

A possible mechanism worth considering is how Ekman pumping drives or regulates the circulation.  The standard paradigm assumes the boundary layer is passive to free-tropospheric ``driving'' mechanisms; Ekman pumping must adjust to provide continuity of mass flux along the bottom edge of the Hadley cell.  However, causality is impossible to determine in balance dynamics, and therefore Ekman pumping should be considered a candidate ``driving'' mechanism.  For instance, a novel picture of Hadley cell dynamics emerges that may apply in the $Ro=-\bar\zeta/f\sim 1$ limit.  Heuristically, the chain of logic of the adjustment to equilibrium is as follows:  1) the mean circulation adjusts the relative vorticity; 2) the omega governor fixes vertical velocities; 3) the width adjusts so that $\bar{\zeta}\sim -f_o$ (making $Ro_L\sim1)$; 4) the Hadley cell strength adjusts; and 5) $\bar{\zeta}$ adjusts to the circulation, and the adjustment loop repeats.  Exactly how this mechanism, if operable in the $Ro\sim 1$ limit, matches scalings with $\Omega^*$ in the $Ro<<1$ limit is unclear. Further progress requires an analysis of the effects of different model boundary layer parameterizations on Ekman pumping and suction, but we leave this to future work.

\subsection{Hadley cell behavior in simulations exhibiting \(Ro\sim1\)}
Figure \ref{fig:caltech} makes clear that large-$Ro$ simulations do not simply follow the scalings predicted by axisymmetric, inviscid theory.  The scalings in all three models are inconsistent with axisymmetric, nearly inviscid theory, which predicts in the small-angle approximation that the strength and width scale separately as $\Omega^{-3}$ and $\Omega^{-1}$ \citep{Held_Hou80}.  In fact, axisymmetric scaling would imply $\omega\sim\Omega^{-2}$, not the constant vertical velocities obtained in our dry simulations.  
Instead, we speculate that Ekman pumping drives vertical velocities that by continuity are limited by the omega governor, and the circulation arranges the vorticity to meet the two constraints, the details of which are determined by the boundary layer friction parametarization.  
As the convective-relaxation and aquaplanet models make clear, changes in $S/f$ are typically offset by changes in $Ro$ in a coordinated fashion such that $\Psi_\mathrm{max}$ has a single, shallow power law.  This qualitative behavior is also apparent in non-eddy-permitting simulations with imposed eddy stresses \citep{Singh_Kuang16}, which could indicate that the circulation adjusts to changes in eddy stresses rather than boundary layer Ekman pumping.

\subsection{Conclusions}
We have presented a new perspective on the Hadley cell descending branch width that relies on vertical velocities in the Hadley cell downdraft being roughly independent of rotation rate.  In the $Ro\ll 1$ limit, the Hadley cell mass flux must provide for the momentum demand of breaking extratropical eddies, and constant vertical velocities dictate that the width and the strength scale with the bulk eddy stresses divided by the planetary vorticity, $S/f$.  The latter quantity scales relatively weakly with rotation rate as compared to predictions of axisymmetric, inviscid theory \citep{Held_Hou80}, and as a result so do the Hadley cell downdraft width and strength.  This theory was tested in simulations with three GCMs over two orders of magnitude in rotation rate.  Single-power-law scalings for Hadley cell widths and strengths are present in all models --- weaker than predicted in axisymmetric theory --- despite displaying a wide and varying range of $Ro$.

Our scaling works well in simulations with Held-Suarez-like forcing. Vertical velocities in the downdraft are constant, which is consistent with the near-constant effective (diabatic plus eddy) heating rate and lapse rate in these simulations.  The Hadley cell downdraft width and strength have the same rotation rate scaling as $S/f$ as our omega governor theory predicts.  Notably, $Ro<1$ for all rotation rates but it is not necessarily small; despite this our scaling holds.

In analogous simulations in an idealized, dry GCM that includes a convective relaxation parameterization, $S/f$ has a non-monotonic scaling with rotation rate, with a negative power law at large rotation rates and a positive one at small rotation rates.  Despite this, the width and the strength follow a single power law in rotation rate, because at weak rotation rates the positive power law in $S/f$ is compensated by an increase in $Ro$.  This $Ro$-number compensation is currently not well understood.  One possibility is that as the Hadley circulation strengthens and widens with decreasing $\Omega^*$, it becomes shielded from the influence of extratropical eddies in a manner similar to the monsoonal circulations \citep{Bordoni_Schneider08, Schneider_Bordoni08}. We also outlined a novel possibility, namely that omega-governed vertical velocities coordinate with Ekman suction in the downdraft region to adjust the cell width and vorticity profiles; importantly, these twin constraints do not depend on the value of $Ro$.  This may help account for how the cell widens and strengthens as in the Held-Suarez-like case, but it also becomes appreciably less eddy-driven (as measured by the bulk Rossby number spanning $Ro\ll1$ to $Ro\sim 1$).  Regardless of $Ro$, downdraft velocities and the effective heating are nearly independent of rotation rate, and as such the omega governor applies across the entire range of simulations, ensuring that the downdraft width and strength share the same scaling across the entire range.

In an idealized moist GCM, our omega governor theory breaks down because $Ro$ and vertical velocities in the downdraft vary significantly with rotation rate.  Eddy stresses and downdraft widths are somewhat weakly dependent on rotation rate. The strength has a considerably steeper power law.  However, the value of $Ro$ systematically varies with rotation rate simulations, and accounting for the dependence of $Ro$ on rotation rate gives a consistent scaling according to (\ref{eq:psimax}) provided $Ro<1$.  Thus, despite the omega governor being invalid for the moist convective model forcing case, the theory provides a good empirical fit to the width and strength given the vertical velocities, eddy stresses, and $Ro$.  Furthermore, the model scalings are consistent with the Ekman constraints outlined above.

Notably, all three model configurations have somewhat similar scalings for the downdraft width, $\sim\Omega^{*-1/3}$, which may suggest a universal driving mechanism.  Consideration of Ekman balance indicates that the omega governor may coordinate with Ekman suction in the downdraft region by adjusting the background vorticity profile, likely by changing the Hadley cell width.  Importantly, this Ekman-omega-governor model of the Hadley cell is independent of the value of $Ro$, unlike other theories.  A detailed accounting of these effects as well as model scalings with other external parameters (planetary radius, equator-to-pole temperature/insolation gradient, etc.) will be explored in future work.

\acknowledgments
J.L.M. acknowledges funding from the Climate and Large-scale Dynamics program of the NSF, award \#1912673, and the University of California Santa Barbara's Earth Research Institute for hosting a sabbatical stay.  S.A.H. acknowledges funding at different stages of this work from NSF AGS Postdoctoral Research Fellowship \#1624740, the Caltech Foster and Coco Stanback Postdoctoral Fellowship, and the Columbia University Earth Institute Postdoctoral Fellowship.

\end{document}